# Phonon anomalies and dynamic stripes

D. Reznik

Department of Physics, University of Colorado, Boulder, CO 80304

**Stripe order where electrons self-organize into alternating periodic charge-rich and magnetically-ordered charge-poor parallel lines was proposed as a way of optimizing the kinetic energy of holes in a doped Mott insulator. Static stripes detected as extra peaks in diffraction patterns, appear in a number of oxide perovskites as well as some other systems. The more controversial dynamic stripes, which are not detectable by diffraction, may be universally present in copper oxide superconductors. Thus it is important to learn how to detect dynamic stripes as well as to understand their influence on electronic properties. This review article focuses on lattice vibrations (phonons) that might show signatures of the charge component of dynamic stripes. The first part of the article describes recent progress in learning about how the phonon signatures of different types of electronic charge fluctuations including stripes can be distinguished from purely structural instabilities and from each other. Then I will focus on the evidence for dynamic stripes in the phonon spectra of copper oxide superconductors.**

## 1. General considerations

Electrons in a doped Mott insulator sometimes form stripe order, consisting of alternating charge-rich lines and charge-poor antiferromagnetic regions. It was predicted theoretically in 1989 [1,2] and discovered in a nickel oxide perovskite in 1994 by neutron diffraction. [3] Static stripe order was subsequently found in some copper oxide perovskites. [4] Diffraction measurements of copper oxide high temperature superconductors do not show magnetic and charge Bragg peaks characteristic of static stripes. However, it was proposed that dynamic stripes, which are much harder to observe than the static ones, [5] play an important role in the copper oxide superconductors. [6,7,8,9]

Periodic charge density modulation, which forms as a result of stripe formation, induces a periodic distortion of the crystal lattice with the same propagation vector. Its magnitude is proportional to the electron-phonon coupling strength. Such distortions can be observed by neutron diffraction as well as scanning tunneling microscopy. [5] Charge density fluctuations associated with dynamic stripes may soften and/or broaden certain phonons signaling an incipient lattice instability. However, structural distortions as well as soft phonons may appear not only because of stripe formation, but also for other reasons. Thus, in order to use phonons as a tool to study dynamic stripes, it is crucial to learn how to distinguish between different origins of soft phonons.

Structural phase transitions can result from purely structural instabilities or from atomic positions reacting to charge density modulation associated with electronic instabilities (such as the tendency to form stripes). In both cases, the phase transition occurs due to a small energy difference between the ground-state structure of low symmetry and a high symmetry structure favored at high temperatures. The high temperature phase is often characterized by a soft phonon mode whose polarization is the same as the "frozen" lattice distortion of the low-T phase. The softening of this mode is most pronounced at the structural transition temperature. In the low temperature phase, the number of phonon branches increases due to the larger unit cell. This phenomenon is called branch folding.

Diffraction is a direct way of detecting structural phase transitions. Static stripes show up in neutron diffraction spectra as extra lattice/magnetic peaks corresponding to charge/spin stripes respectively. Many solids do not undergo structural phase transitions, but exhibit incipient electronic and/or lattice instabilities, which are undetectable by diffraction. In these cases, softening and broadening of phonons (which is often temperature-dependent) provides a unique window on competing phases, which often play an important role in many phenomena including stripe formation.

My goal here is to review recent progress in studies of phonons in correlated electron systems with the focus on phonon anomalies that may be associated with the formation of charge stripes. This article is organized as follows. First I will discuss soft phonons in different systems and demonstrate that, although phonon behavior is very similar near different types of lattice instabilities, it may be possible distinguish between them. Then, I will focus on the features of soft phonons that may be specific to dynamic stripes.

This article partially overlaps and complements two other recent reviews. [10,11]

## 2. Different types of structural and electronic instabilities.
Crystal structure as well as frequencies of atomic vibrations are ultimately determined by interactions between all atomic nuclei and all electrons in the crystal, so in this sense all lattice instabilities are driven by electron-phonon coupling. However, one can distinguish between the instabilities caused by electron-phonon coupling directly (which I call structural instabilities), and the instabilities where the electronic state itself undergoes a phase transiton, which then pulls the lattice along (which I call electronic instabilities). Stripe formation is an example of the latter.

Phonon measurements can provide insights into mechanisms underlying second order displacive structural transitions that involve soft phonon behavior. In these transitions atomic displacements in a crystal change bond lengths and/or angles, without severing the primary bonds. Such transitions are associated with soft phonons whose eigenvectors are close to the character of the atomic displacements that take place during the transition. The contribution to the Hamiltonian from the soft phonon with the generalized coordinate X, H(X), can always be approximated by adding a double-well potential term to the regular harmonic potential. It is useful to approximate the bottom of the double well potential as:

$V_{double-well}(X) = aX^2 + bX^4$, where $b > 0$.

If $a < 0$, the two sides of the well with the minima at $X = \pm\sqrt{-a/2b}$ represent the lattice distortion in one or another symmetry-equivalent direction. At temperatures higher than the barrier between the potential wells, ($k_BT > a^2/4b$), the average structure is undistorted, i.e, $<X> = 0$. But the anharmonic potential makes the phonon soft and broad. This broadening and softening becomes enhanced on cooling towards the transition temperature, as the amplitude of the vibration decreases and the phonon feels the anharmonic bottom of the potential more and more. Below $T_c$, the lattice settles at the minimum of one of the wells with $<X> = +\sqrt{-a/2b}$ or $<X> = -\sqrt{-a/2b}$, and the phonon hardens and narrows.

If $a > 0$, the phase transition will not occur, but soft phonons induced by anharmonicity and/or electron-phonon coupling may be observed.

Usually (but not always) the phonon spectral function is that of a damped harmonic oscillator. In the limit of weak damping it is close to a Lorentzian centered at the phonon frequency whose linewidth is proportional to the inverse phonon lifetime.

This behavior is common to all second order real or incipient displacive transitions regardless of their origin (see sec. 2.1, 2.2). It is possible to distinguish between different mechanisms behind them (e.g. structural vs. electronic) only by detailed observations, calculations, and analysis focusing on subtle features of the phonon spectra. Without detailed understanding of soft phonons in general, little can be learned about dynamic charge stripes from phonon measurements on the copper oxides.

Sec. 3 provides highlights of substantial recent advances in understanding of how structural and electronic instabilities affect phonons in real systems. This information is then used in sections 4 and 5 to discuss experimentally observed phonon anomalies in copper oxides and their possible relationship to dynamic stripes.

## 2.1 Structural instabilities

Structural instabilities can occur when an anharmonic double-well potential for the anomalous phonon is induced by a mismatch of bond lengths, which favors a small distortion away from a high-symmetry structure. We will discuss an example of such an instability: the HTT-LTO transition in $La_{2-x}Sr_xCuO_4$. (Sec. 3.1)

In metals, strong coupling of conduction electrons to phonons can also induce a lattice deformation, which is usually discussed in terms of enhanced coupling of a partcular phonon to electron-hole excitations of conduction electrons across the Fermi surface. These phonon anomalies can be better understood in terms of phonon renormalization by conduction electrons. (Sec. 3.3,3.4,3.5,3.6)

## 2.2 Electronic instabilities

Phonon anomalies associated with electronic instabilities are induced by screening of the phonon by charge density fluctuations associated with the instability. Electronic instabilities can be of two types. One occurs when the Fermi surface nesting induces a charge density wave (CDW) or a spin density wave (SDW) instability. (Sec. 3.2,3.6,3.7) The second one originates from electronic correlations such as the stripe-ordering instability.

## 3. Phonon anomalies in model systems
## 3.1 An example of a purely structural instability without the participation of conduction electrons: HTT-LTO transition in $La_{2-x}Sr_xCuO_4$.

A good example of a purely structural instability is the transition from the high temperature tetragonal (HTT) to the low temperature orthorhombic (LTO) phase in $La_{2-x}Sr_xCuO_4$. It occurs due to the mismatch of La-O and Cu-O bond lengths. This mismatch introduces strain into the tetragonal structure, which is stable at high temperatures. In the LTO phase, a small rotation of the $CuO_6$ octahedra around the 110 direction (defined based on the HTT unit cell) relieves the strain. Braden et al. [12] investigated the HTT-LTO transition in $La_{1.87}Sr_{0.13}CuO_4$ in detail. In particular, they found soft phonon behavior: the phonon that corresponds to the rotations of the octahedra has a pronounced minimum in its dispersion at the wavevector where the additional Bragg peak appears in the LTO phase. (see Fig. 1) This dispersion dip is the largest at the phase transition temperature, $T_{LTO}$.

At temperatures higher than the barrier between the wells of the double-well potential discussed in Sec. 2, ($k_BT > a^2/4b$), the average structure is tetragonal with zero octahedral tilt, but the anharmonic potential already softens and broadens the rotational mode of the $CuO_6$ octahedra. This broadening and softening becomes enhanced on cooling towards the transition temperature, as the amplitude of the vibration decreases and the phonon increasingly feels the anharmonic bottom of the potential.

One can exclude electron-phonon coupling to conduction electrons from the mechanism of the transition, because the same instability occurs in the undoped $La_2CuO_4$, which is an insulator. Furhtermore, the metal-insulator transition that occurs with Sr doping has no observable impact on $T_{LTO}$.

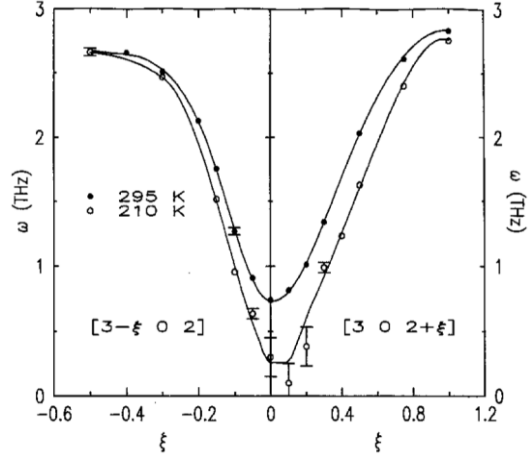

Figure 1. The dispersion of the soft mode in $La_{1.67}Sr_{0.13}CuO_4$ near the structural phase transition (210K) and at room temperature (from [12]).

### 3.2 Kohn anomalies

Phonons behave very similarly when the structural distortion results from an electronic instability although the underlying physics is quite different.

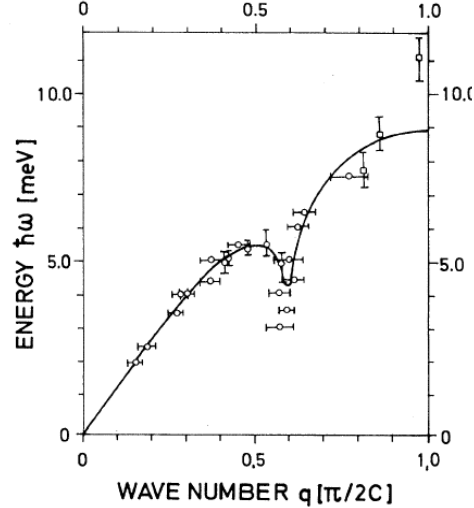

Figure 2. LA phonon branch of $K_2Pt(CN)_4Br_{0.30} \cdot 3D_2O$ at room temperature. Solid line represents the result of a calculation based on the simple free-electron model as discussed in Ref. [14].

In metals with Fermi surfaces, phonons may couple to singularities in the electronic density of states, which appear at specific wavevectors. Here the phonon renormalization associated with the real or incipient phase transition is discussed in terms of electron-phonon coupling. The coupling is between the phonon and the low-energy two-particle electronic response. Singularities in the electronic response can produce sharp features in phonon dispersions called Kohn anomalies. [13] These typically become stronger with reduced temperature, due to the sharpening of the Fermi surface. A classic example of this behavior is one-dimensional conductors such as $K_2Pt(CN)_4Br_{0.30} \cdot 3D_2O$ (KCP). Due to the one-dimensionality of the electronic states, these systems are characterized by the Fermi surface nesting at $2k_f$ ($\hbar k_f$ is Fermi momentum). This nesting greatly enhances the number of possible electronic transitions at $2k_f$ compared to other wavevectors, which results in softer and broader phonons. For this reason acoustic phonons in KCP show pronounced dips at $\mathbf{q}=2k_f$ ($\mathbf{Q}$ is the total wavevecor, $\mathbf{q}$ is the reduced wavevector.) [14] (see Fig. 2)

The amount of this phonon softening and broadening depends on the details of the interaction and varies greatly between different systems with Kohn anomalies. Often the broadening is much smaller than the experimental resolution, thus only the softening appears in the experiment.

### 3.3 ab-initio calculations and the role of the q-dependence of the electron-phonon matrix element.

Electron-phonon scattering as well as anharmonicity influence the phonon frequency and reduce the lifetime (real/imaginary parts of the phonon self-energy).

Isolating the role of electrons near the Fermi surface requires knowing phonon frequencies and linewidths in the absence of coupling of phonons to electrons near the Fermi surface. Determining these accurately is typically a challenge. The simplest way to model phonons is by balls-and-springs models with the atomic nuclei serving as balls and the Coulomb forces screened by the electrons as springs. Shell models are more sophisticated modifications of this approach. Including only short-range interactions gives smooth phonon dispersions. Such models can be fit to the experimental dispersions that do not contain any sharp dips. Then deviations from these dispersions may indicate an incipient electronic or lattice instability. However, this phenomenological approach by itself is unable to distinguish between different types of instabilities discussed above.

In fact even defining the real part of the phonon self-energy, is not at all straightforward, because the unrenormalized dispersions are not easy to determine. Thus very precise measurements of phonon frequencies and linewidths do not, in a general case, provide direct information on electron-phonon coupling. However, combining careful measurements with detailed calculations and general arguments, can often elucidate the mechanism of the phonon anomalies.

It was noticed early on that Fermi surface nesting alone cannot adequately explain phonon softening resembling Kohn anomalies in many cases. For example, dispersions of certain phonons in NbC and TaC dip strongly at wavevectors where the FS nesting is relatively weak. (see Fig. 3) [15,16] These could be modeled by including either very long-range repulsive interactions or by adding an extra shell with attractive interactions. [16] S.N. Sinha and B.N. Harmon [17] introduced **q**-dependent electron-phonon coupling to explain these effects.

They included **q**-dependent dielectric screening into their model and obtained good agreement with experiment for certain values of adjustable parameters.

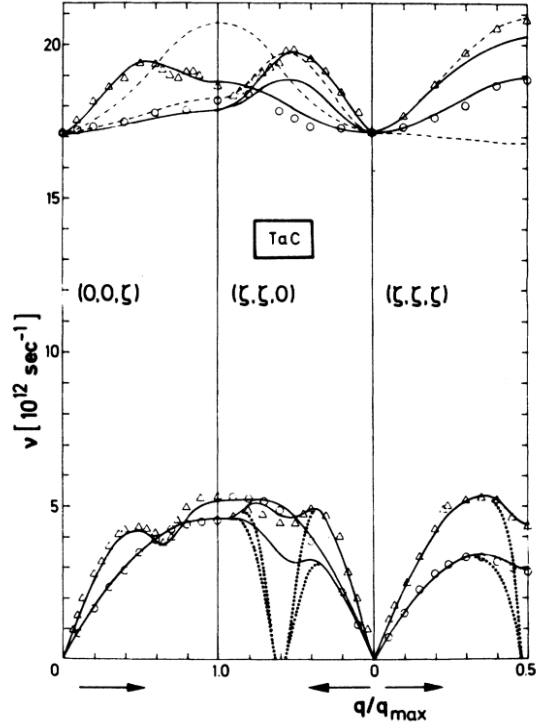

Figure 3. Phonon dispersions of TaC (data points) [15] and calculations based on the double- shell model of W. Weber. [16] Note that the dispersion dip at $\mathbf{Q}=(0,0,0.6)$ originates mostly from the $\mathbf{q}$-dependence of electron-phonon coupling strength.

Further theoretical development led to ab-initio calculations based on the density functional theory (DFT) in the local density approximation (LDA) or in the generalized gradient approximation (GGA). These treat the electronic correlations at the mean field level and can very accurately predict phonon dispersions and linewidths in systems with structural instabilities as well as the electronic instabilities related to Fermi surface nesting. [18] However, it is very difficult to distinguish between these two scenarios and isolate contributions of specific electronic states to the phonon self-energy using this approach. Since it is necessary to go beyond the approximations built into LDA and GGA to obtain stripes, any phonon anomaly that is reproduced in these approximations cannot originate from interactions between phonons and stripes.

R. Heid et al. made an attempt to overcome this problem in Ru. [19] First they performed both the ab-initio DFT/LDA calculations and detailed measurements on a high quality single crystal. The phonon dispersions in Ru had a pronounced softening near the M-point, which was well reproduced by theory. The calculated softening became much weaker when electron-phonon coupling was made $\mathbf{q}$-independent. (see Fig. 4) This result indicated that both the Fermi surface nesting and the $\mathbf{q}$-dependence of electron-phonon matrix element were important (the latter more than the former). The phonon dispersion dip disappeared entirely upon the exclusion of conduction electrons from the calculation. In addition, the entire dispersion hardened substantially, which may be an artifact of the procedure used to leave out the conduction electrons. This method showed that the phonon dispersion dips in Ru originate from coupling to conduction electrons, but the bare dispersions obtained by excluding them were somewhat arbitrary. This study illustrates the difficulty in separating the bare dispersion from the real part of the phonon self-energy. It is not important for reproducing or predicting phonon anomalies, but is important for making connections with other experiments and for identifying the origin of the phonon dips.

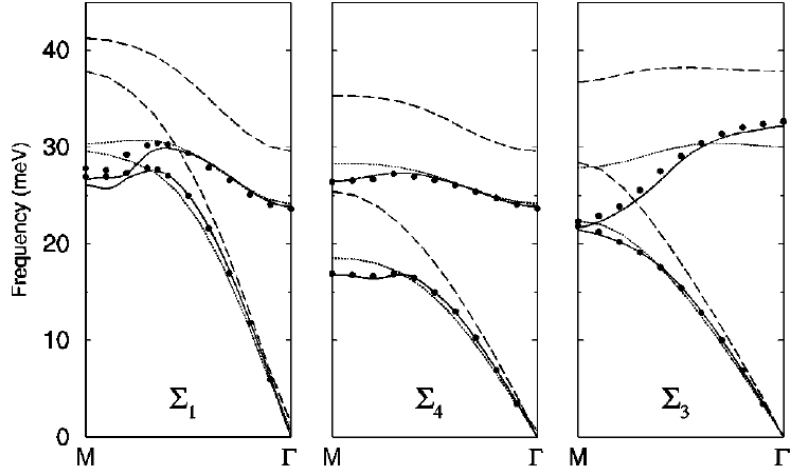

Figure 4. Phonon dispersions in Ru. (from [19]) The data points correspond to experimental results. The solid lines represent results of the full DFT/LDA calculation. Dotted lines were obtained by replacing **q**-dependent electron-phonon matrix element with the average value (which was **q**-independent). The dashed lines represent the DFT calculation without including the conduction electrons. Disappearance of the phonon dispersion dip near the M-point upon exclusion of conduction electrons demonstrates that the dip results from electron-phonon coupling.

### 3.4. A reinterpretation of the origin of the CDW formation in NbSe$_2$ based on phonon measurements.

NbSe$_2$ has been considered as a classic quasi-2D CDW system with an incommensurate CDW induced by Fermi surface nesting. However, based on theoretical arguments, it has been suggested that the formation of an incommensurate superstructure in this material results from the **q**-dependence of electron-phonon coupling, not Fermi surface nesting, i.e. that it is not a true CDW but is a structural instability. [20,21,22] However, these ideas have not been tested by neutron scattering measurements of phonons, because the available samples were too small for a comprehensive temperature-dependent study.

Recently F. Weber et al. performed detailed measurements of phonon dispersions using inelastic x-ray scattering (IXS), which does not require large samples. [23] Figure 5 shows the phonon dispersion in NbSe$_2$ as a function of temperature with the phase transition occurring at 33K. The **q**-dependence of the phonon softening is in marked contrast to the sharp, cusplike dips that normally characterize Kohn anomalies at $2k_F$ due to the Fermi surface nesting. In 2H-NbSe$_2$ the phonon renormalization extends over 0.36Å$^{-1}$, or over half the Brillouin zone, and the critically damped region (where the phonon frequency goes to zero) extends over 0.09 Å$^{-1}$, which is much broader than the experimental **q**-resolution. This behavior clearly rules out a singularity in the electronic response in 2H-NbSe$_2$.

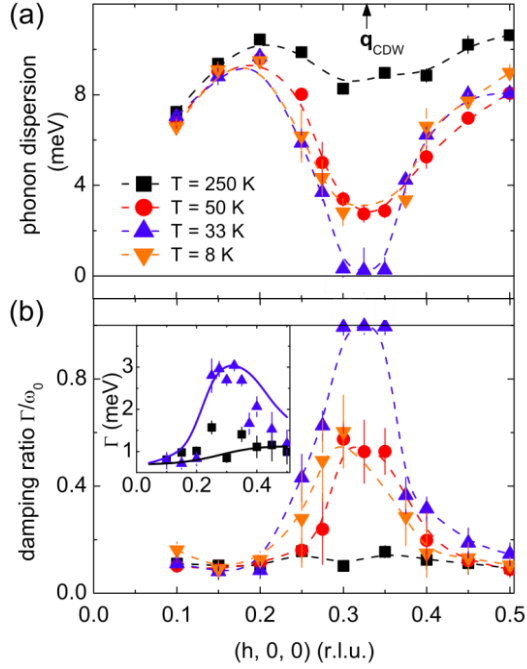

Figure 5. Experimentally obtained dispersion and damping ratio of the soft-phonon branch in 2H-NbSe$_2$ at four temperatures 8K<T<250K. Plotted are (a) the frequency of the damped harmonic oscillator $\omega_q = \sqrt{\tilde{\omega}_q^2 - \Gamma^2}$ and (b) the damping ratio $\Gamma/\omega_q$. Lines are guides to the eye. Note that phonons at h = 0.325; 0.35 and T = 8 K were not detectable due to strong elastic intensities. The inset in (b) shows the experimentally observed damping of the damped harmonic oscillator (symbols) and scaled DFT calculations with σ = 0.1 eV (blue) and 1 eV (black). (from Ref. [23])

By detailed comparison with the DFT/LDA calculations, Weber et al. showed that the structural instability results from the **q**-dependence of the electron-phonon matrix element. They smeared the calculated Fermi surface by different amounts and found a strong effect of smearing on the phonon dispersion, which suggests that the interaction with conduction electrons drives the instability.

The observed phonon dispersion result is in contrast with the phonon anomaly reported in a quasi1-D metal ZrTe$_3$, [24] where the dispersion dip occurs over a much narrower region of reciprocal space, which is similar in size to KCP.

F. Weber et al concluded that the CDW wavevector is determined by the wave vector dependence of the electron-phonon coupling, as proposed in [20,21,22], not by the position of the broad maximum of the two-particle electronic response measured by ARPES [25].

### 3.5 Phonon anomalies in conventional superconductors
There are two interesting electron-phonon effects observed in measurements of phonon dispersions and linewidths in conventional superconductors. One is the normal state anomalies whose study can elucidate, which phonons contribute the most to the mechanism of superconductivity. The other is the effect of the superconducting gap 2Δ on the phonon spectra below T$_c$, which can be used as a probe of the superconducting gap.

In MgB$_2$ the high superconducting transition temperature, T$_c$, is explained by strong electron-phonon coupling of E$_g$ modes around 80meV near the zone center. A strong dip in the dispersion of these phonons observed in experiments and reproduced by LDA calculations is a clear signature of this coupling. [26,27]

The ab-initio calculations based on DFT/LDA can predict all physical properties of materials that depend on electronic band structure, phonon dispersions, and electron-phonon coupling without adjustable parameters. In particular, they can be used to calculate $T_c$ based on Migdal-Eliashberg theory. [28] Several ab-initio calculations of phonon dispersions as well as $T_c$s were performed for the transition metal carbides and nitrides in order to explain relatively high transition temperatures in some and not in others. [29,30,31] They correctly reproduced phonon dispersions including the anomalies discussed in Sec. 3.3, [29,31,32] and established a correlation between the phonon anomalies and $T_c$. They also suggested that the phonon anomalies are associated with the Fermi surface nesting [31], but more detailed calculations along the lines of Ref. [19] need to be performed to separate the role of nesting from the enhancement of the electron-phonon matrix element. One interesting possibility that may need to be explored, is that additional screening near the nesting wavector may enhance the electron-phonon matrix elements, thus the Kohn anomalies at the nesting wavevectors may be stronger than expected from enhanced electronic response due to nesting alone.

In a conventional superconductor with $T_c$=15K, YNi$_2$B$_2$C, the transverse acoustic phonons near $\mathbf{q}$=(0.5 0 0) soften and broaden on cooling [33] as expected from electron-phonon coupling. W. Reichardt et al. calculated phonon frequencies and linewidths using LDA. [34] These calculations reproduced this phonon anomaly and correctly predicted an additional wavevector ($\mathbf{q}$=0.5 0.5 0) where acoustic phonons couple strongly to conduction electrons. (see Fig. 6) [35] The energies of both soft phonons are comparable to the superconducting gap energy. In this case the opening of the superconducting gap has a strong influence on the phonon spectral function.

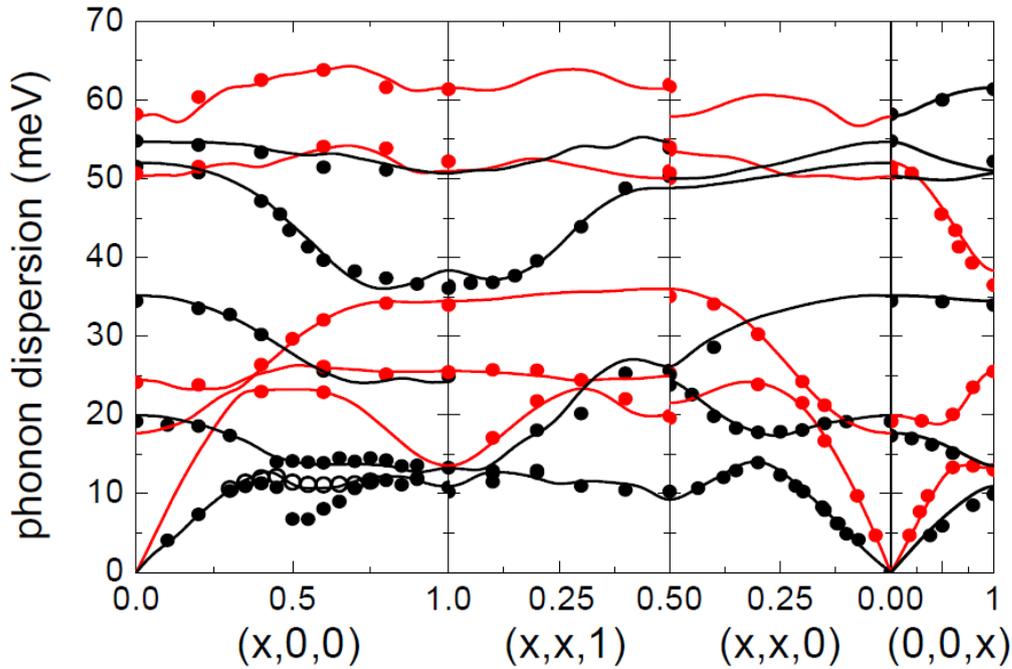

Figure 6. Calculated (lines) and observed (filled dots) phonon frequencies in YNi$_2$B$_2$C at 20K. Open dots along [100] were measured at 300 K. Branches shown in red/black refer to phonons of predominantly longitudinal/transverse polarization respectively. The horizontal axes denote different crystallographic directions in reciprocal lattice units (r.l.u.). The theoretical results were scaled up by a factor of 1.03 (from Ref. [34]).

These phonons were so broad, that their normal state spectra extended below the low temperature superconducting gap $2\Delta$. In this case, when the $2\Delta$ opens in the electronic spectrum in the superconducting state, the damped harmonic oscillator approximation of the phonon breaks down. P.B. Allen et al. [36] developed a theory precisely for this case, which predicted that phonon lineshapes in the superconducting state should contain either a step or a sharp peak very close to 2Δ depending on the values of the phonon energy, electron-phonon coupling and 2Δ. Normal state lineshape fixes the first two parameters, which leaves only one adjustable parameter, i.e. the superconducting gap. Detailed measurements of F. Weber et al. [37] confirmed this theory. (Fig. 7) They also suggested how to use phonon measurements to determine the magnitude of 2Δ and to probe gap anisotropy.

The above analysis allows a reliable determination of the origin of phonon renormalization in YNi$_2$B$_2$C. Since the superconducting gap affects the phonon lineshapes as predicted by P.B. Allen et al., the phonon renormalization originates from the interaction between conduction electrons and phonons. Breadth of the phonon anomalies in **q**-space argues against the nesting scenario: the **q**-width of the anomalous region is closer to the one in NbSe$_2$ than to the 1D metals. Furthermore, the value of the superconducting gap extracted from phonon lineshapes was different for different wavevectors, which proves that soft phonons with different **q** connect different parts of the Fermi surface. Thus the phonon anomaly originates primarily from the **q**-dependence of electron-phonon coupling rather than from the nesting of the Fermi surface. LDA calculations predict phonon softening at the correct wavevectors (with the caveat that the softening is more pronounced at q=(0.5-0.7, 0, 8) at low T, so it is not necessary to invoke electronic correlations to explain experimental results.

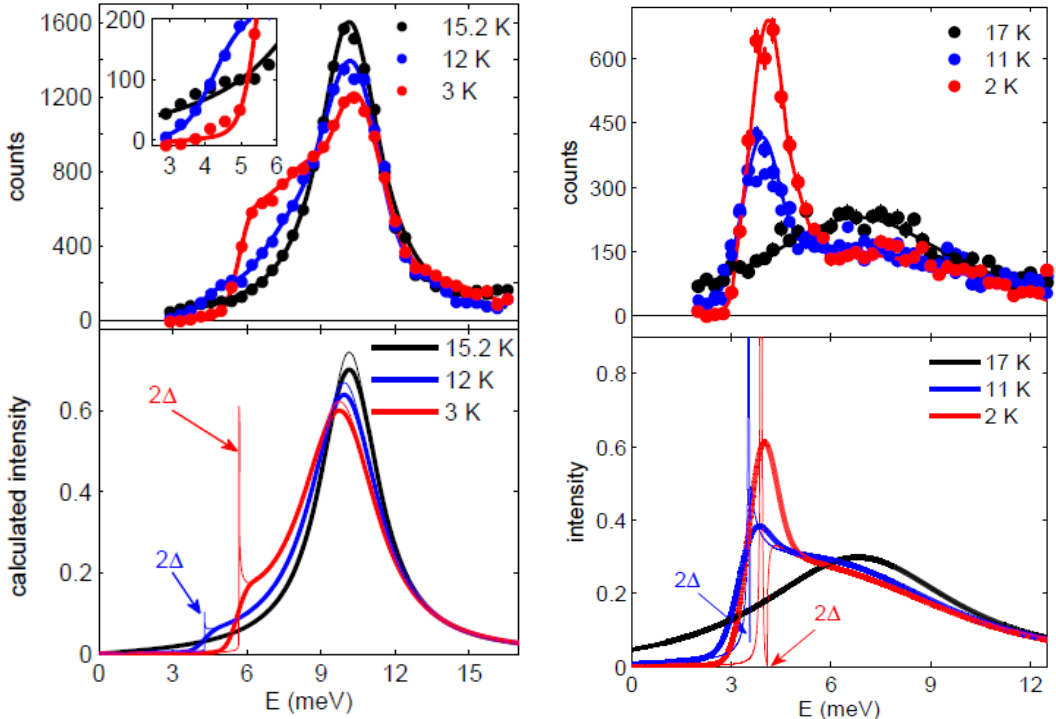

Figure 7. Top panels: Evolution of the neutron-scattering profile measured on YNi$_2$B$_2$C at **Q** =(0.5,0.5,7) (left) and **Q** =(0.5,0.8) (right) above and below T$_c$=15K. Bottom panels: Thin lines represent calculated phonon lineshapes based on the theory of Allen et al. [36], using parameters extracted from the lineshape observed in the normal state. The thick lines are obtained after the convolution of the calculated lineshape with the experimental resolution. (from Ref. [37])

Similar results were obtained on subsequent phonon measurements of Nb where phonon lineshapes were in almost perfect agreement with conventional theory of P.B. Allen et al. discussed above, [38] although evidence for an effect of correlations on phonon linewidths beyond this theory in Nb and Pb was reported in [39] based on ultra-high resolution neutron scattering measurements.

### 3.6 Phonon anomalies with and without the Fermi surface nesting in Chromium

A different situation appears in Chromium where the Fermi surface nesting is responsible for an incommensurate spin density wave (SDW) at a nesting wavevector $\mathbf{q}_{sdw}$=(0.94,0,0) and, as a secondary effect, of the charge density wave (CDW) at $\mathbf{q}_{cdw}$=(0.11,0,0). [40] W. M. Shaw and L. D. Muhlestein measured phonon dispersions in Chromium by INS. [41] They reported soft phonons around $\mathbf{q}$=(0.9,0,0), which is near $\mathbf{q}_{sdw}$, as well as near $\mathbf{q}$=(0.45,0.45,0) where some nesting has also been calculated. However, the neutron data were not accurate enough to establish their exact wavevectors.

Recently Lamago et al. [42] performed a more precise and comprehensive set of measurements by IXS, which showed that the two anomalies actually appear at $\mathbf{q}_{sdw}$ and $\mathbf{q}$=(0.5,0.5,0) respectively. A surprising result was that a transverse phonon branch softened throughout the zone boundary between $\mathbf{q}$=(0.5, 0.5, 0) and (1, 0, 0), i.e. for $\mathbf{q}$=(0.5+h,0.5-h,0), where 0<h<0.5. LDA-based calculations performed as a part of this investigation, successfully reproduced the observed phonon softening. (Fig. 8) However, the electronic response function that couples to the phonons obtained from the same calculations showed no clear features corresponding to the phonon dips along the zone boundary. Thus these phonon dips come exclusively from the $\mathbf{q}$-dependence of the electron-phonon coupling strength.

This result was further corroborated by the effect of the numerical smearing of the Fermi surface on the calculated phonon dispersions. Such a smearing makes it possible to isolate coupling to electronic excitations near the Fermi surface. If it has a strong effect on the calculated phonon dispersions, electrons near the Fermi surface contribute significantly to the phonon self-energy. Otherwise, the coupling to these electrons is small.

In Cr this smearing suppressed only the calculated effect at $\mathbf{q}$=(0.94,0,0). It had no effect at $\mathbf{q}$=(0.5+h,0.5-h,0) for any h including h=0. (Fig. 8a,b). Thus the phonon dispersion dip at $\mathbf{q}$=(0.5+h,0.5-h,0) comes exclusively from an interaction with electrons far from the Fermi surface. This is true even for h=0, which is near a nesting feature previously thought [41] to be responsible for phonon softening. This nesting feature disappears even at small h, but the phonon renormalization does not become smaller in either the calculation or the experiment. In contrast with the earlier work [41], which did not include this type of analysis or measurements for h>0, D. Lamago et al. concluded that this nesting feature makes a negligible contribution to the phonon self-energy. [42].

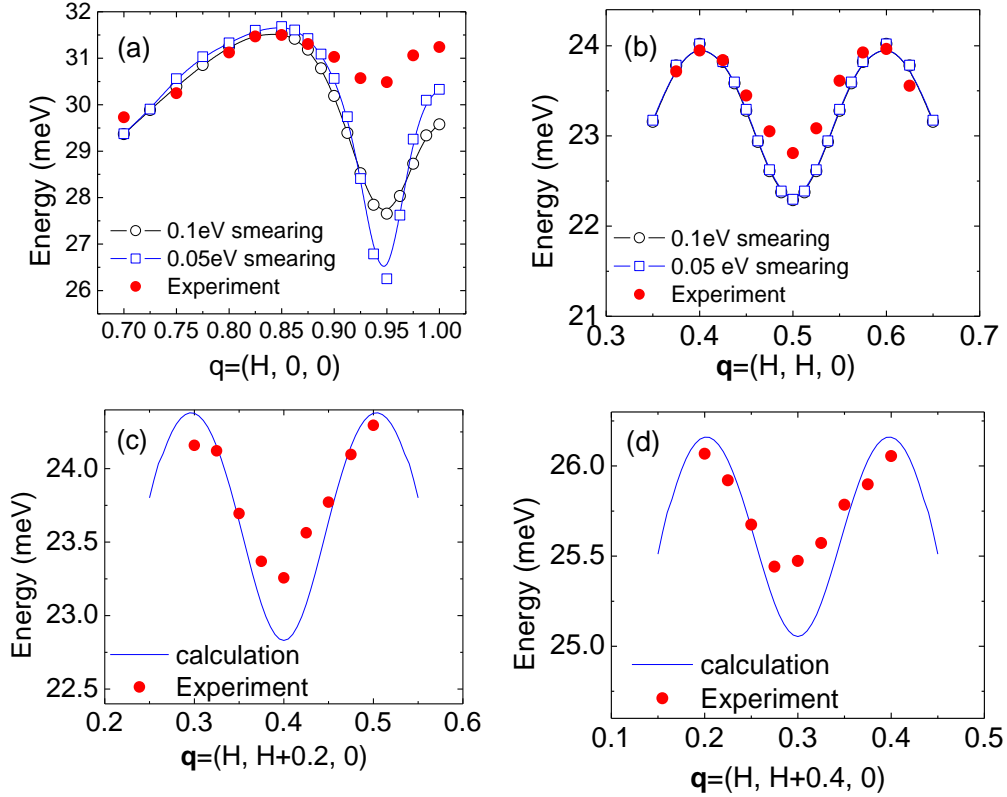

Figure 8. A comparison between measured (red) and calculated (blue, black) phonon dispersions and the LDA calculation in Cr. Only the data for the lowest energy transverse phonons near the zone boundary line connecting $\mathbf{q}=(1,0,0)$ and $\mathbf{q}=(0.5,0.5,0)$ are shown. The smearing has a strong effect in (a) but not in (b). (from Ref. [42])

Thus different types of phonon anomalies can appear in the same material.

### 3.7 Phonon anomalies related to stripes in La$_{1.69}$Sr$_{0.31}$NiO$_4$

Charge modulation due to stripe formation is not directly related to Fermi surface nesting, but results from electronic correlations (see sec. 1). Intense research effort has been devoted to trying to understand the physics of stripes because dynamic stripes may play an important role in copper oxide superconductors. Most work focused on the magnetic component of the stripes, and charge stripes still remain relatively unexplored. In particular, the interaction between charge stripes and phonons is still poorly understood.

According to a model proposed by Kaneshita et al., [43] steeply dispersing Goldstone modes arising from charge stripe order should mix with phonons at the charge stripe ordering vectors and the resulting anticrossing would be observed as phonon line splitting or phonon line softening and/or broadening if the experimental resolution is insufficient. (see Fig. 9)

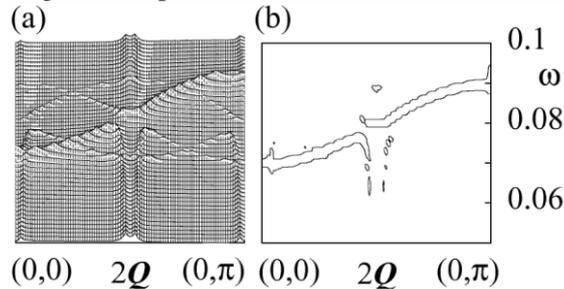

Figure 9. Renormalized phonon spectral functions assuming an interaction between phonons and collective stripe modes. (a) is a logarithmic plot to emphasize the small intensity of the shadow bands. (b) is a contour plot shown for high intensity. The simple sinusoidal form is assumed for the unperturbed phonon. [43]

It is widely believed that charge stripes in the nickelates and cuprates should couple most strongly to Ni-O or Cu-O bond-stretching phonons, hence these phonon branches have been investigated most thoroughly. It is also believed that the L-component (along the direction perpendicular to the Ni-O planes) can be ignored except at low energies and the same effects should be observed at all L. Tranquada et al. searched for the effect of stripes on Ni-O bond stretching phonons in $La_{1.69}Sr_{0.31}NiO_4$. [44] The stripe-related anomaly was expected at $\mathbf{q}$=(0.31,0.31,L), but instead of an anomaly at this wavevector they found that the phonon branch splits on approach towards the zone boundary ($\mathbf{q}$=(0.5,0.5,0) in the 110 direction (in the tetragonal notation). (see Fig. 10) This effect was explained in terms of a toy model containing inhomogeneous distribution of force constants as would occur in the case of static stripes with a large amplitude of charge inhomogeneity.

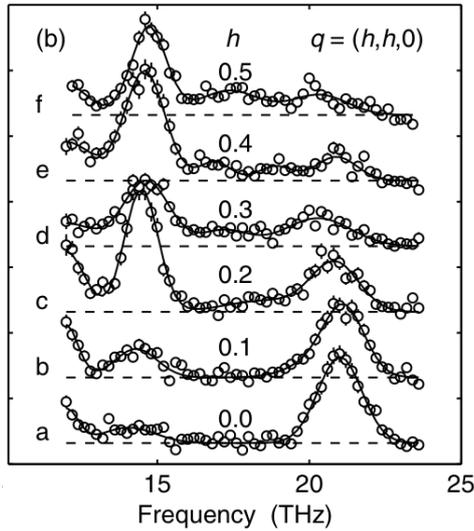

Figure 10. Neutron scattering measurements of phonons in $La_{1.69}Sr_{0.31}NiO_4$ at T ~10 K for $\mathbf{q}$ =(h, h, 0); dashed lines indicate the constant background. Measurements were performed in various equivalent zones, with varying L component, in order to avoid spurious peaks due to accidental Bragg scattering by the sample. (from [44])

A number of important open questions about the interactions between stripes and phonons in the nickelates remain. In particular, acoustic phonon branches have not been investigated. Ref. [44] presented data above the stripe ordering temperature only at the zone center and the zone boundary. Thus it is unknown what happens to phonons at the stripe-ordering wavevectors at temperatures above the ordering temperature where stripes should be purely dynamic. This last question is especially important, because stripes in the cuprate suprconductors are dynamic if they exist.

### 3.8 Lessons learned from model systems

It follows from the investigations reviewed above as well as other similar studies that all types of structural and electronic instabilities have a similar impact on phonon dispersions as well as line-broadening. The following steps are essential to correctly identify the origin of anomalous phonon softening and broadening most of the time:

1. Measurements of phonon dispersions and linewidths
2. Measurements of the temperature dependence of phonons at anomalous wavevectors.
3. DFT calculations of phonon dispersions for different values of the smearing of the Fermi surface.

4. ARPES measuremtns and/or DFT calculations of the electronic joint density of states in either LDA or GGA approximation to determine the nesting properties of the Fermi surface.

*The phonon anomalies predicted by the DFT calculations that appear at the calculated and/or measured nesting wavevectors and are very sharp in q-space* result from the Fermi surface nesting. For example, the Fermi surface nesting-related anomalies in KCP, ZrTe$_3$ and Cr (near $\mathbf{q}$=(0.95,0,0)) cover about 5% of the Brillouin zone or less. In this case, the calculated anomalous phonon energies should be sensitive to the smearing of the Fermi surface.

*The anomalies predicted by DFT calculations that are broad in q* are not associated with nesting, but originate from structural instabilities or q-dependence of electron-phonon coupling. For example, in La$_2$CuO$_4$, Cr (away from $\mathbf{q}$=(0.95,0,0)), and NbSe$_2$ the width of the anomalous phonon regions is about 15% of the Brillouin zone. If the calculated phonon dispersions are sensitive to the smearing of the Fermi surface, these structural instabilities originate from interactions with conduction electrons via an enhanced electron-phonon coupling matrix element. In this case soft phonon lineshapes can be deformed by the opening of a gap in the electronic spectra such as the superconducting gap. If the calculated phonon dispersions are not sensitive to the smearing of the Fermi surface (such as in Cr away from $\mathbf{q}$=(0.95,0,0)) or if there are no conduction electrons present (such as in La$_2$CuO$_4$), renormalization of the anomalous phonons by electrons near the Fermi surface can be ruled out. In this case the relevant physics can be best described in terms of an anharmonic potential or coupling to electronic states far from the Fermi surface.

*Phonon anomalies not explained by the DFT/LDA or DFT/GGA calculations* originate from strong electronic correlations beyond the LDA/GGA (mean field) level such as the tendency to form stripes. The first step to identify such phonon anomalies is to look at features in the phonon spectra that are not predicted by the calculations and have the $\mathbf{q}$-dependence that may be associated with stripe formation. However, this prescription is insufficient to unambiguously identify signatures of dynamic stripes. More work on model systems where stripes are known to exist is necessary to make further progress. The study of J.M. Tranquada et al. [44] was the first step in this direction. Further investigations are under way. [45]

## 4. Phonon anomalies possibly related to stripes in doped La$_2$CuO$_4$
### 4.1 Summary of early work
Calculations performed soon after the discovery of high temperature superconductors suggested that electron-phonon coupling is too weak to account for high temperature superconductivity. [46,47] However, measurements performed in conjunction with shell model calculations showed that the bond-stretching branch softened strongly towards the zone boundary as doping increased from the insulating phase to the superconducting phase. [48] This behavior pointed towards strong electron-phonon coupling and a possible role of the zone boundary "half breathing" bond-stretching mode in the mechanism of high temperature superconductivity. [49] It later became apparent that this trend continues into the overdoped nonsuperconducting phase, which indicates that zone boundary softening is related to the increase of metallicity with doping rather than to the mechanism of superconductrivity (Fig. 11). [50]

These experiments and related calculations are extensively covered in the previous review by L. Pintschovius. [10]

A recent investigation of Park et al. [51] also revealed that the linewidth of the zone boundary bond-stretching phonon (around $\mathbf{q}$=(0.5,0,0)) broadens strongly when the doping, x, is reduced and then narrows abruptly at x=0. They explained this effect in light of NQR/NMR results showing that the doped carrier concentration in the Cu-O planes is inhomogeneous, and since the zone boundary phonon frequency depends on doping, the experimental lineshape broadens. This sensitivity to doping increases towards small x, which results in the increased phonon broadening. The disappearance of the effect at x=0, where there are no doped carriers, also naturally follows from this model.

Here I will focus not on the zone boundary, but half-way to the zone boundary in the [100]-direction (along the Cu-O bond) where the most interesting physics has been observed. This work began with the INS experiments of McQueeney et al. [52] who reported anomalous lineshape and temperature dependence of the bond-stretching phonons in $La_{1.85}Sr_{0.15}CuO_4$ near $\mathbf{q}$=(0.25, 0, 0) (Fig. 12) and interpreted these results in terms of line splitting due to unit cell doubling induced by stripe formation, which is similar to the branch splitting later reported for stripe-ordered $La_{1.69}Sr_{0.31}NiO_4$ by Tranquada et al. [44] (see above). (Fig. 10) It is important to emphasize here that long-range-ordered static stripes do not form in this compound, and the stripes must be dynamic if they exist.

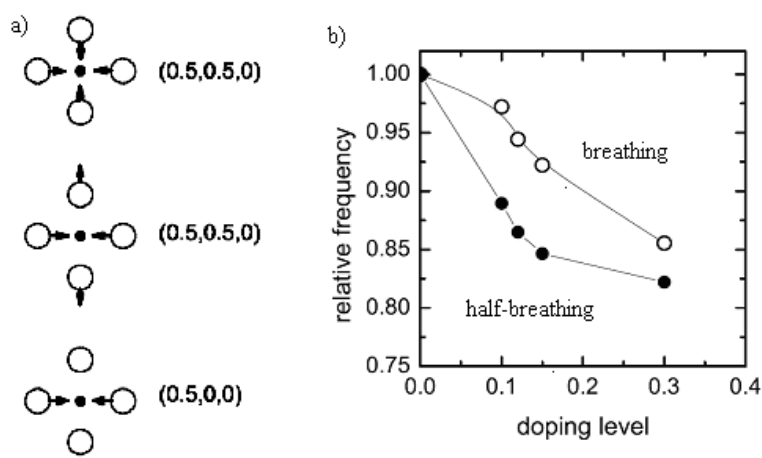

Figure 11. a) Displacement patterns of zone-boundary bond-stretching modes in cuprates. Top: longitudinal mode in the [110]- direction (breathing mode); middle: transverse mode in the [110]-direction (quadrupolar mode); bottom: longitudinal mode in the [100]-direction (half-breathing mode). Circles and full points represent oxygen atoms and copper atoms, respectively. Only the displacements in the Cu-O planes are shown. All other displacements are small for these modes. b) Schematic of doping dependence of the breathing $\mathbf{q}$=(0.5,0.5,0) and half-breathing $\mathbf{q}$=(0.5,0,0) zone boundary mode frequencies. This behavior is probably not directly related to the mechanism of superconductivity, since the softening continues into the overdoped nonsuperconducting part of the phase diagram. (From Ref. [50])

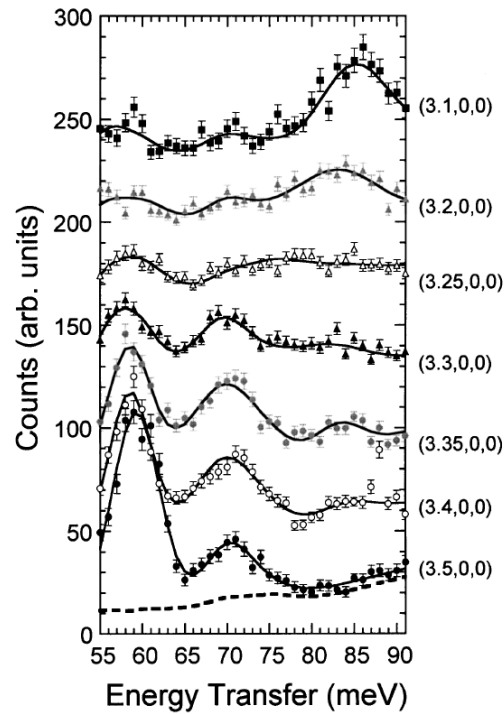

Figure 12. Bond-bending (around 60meV) and bond-stretching (above 65meV) branches in optimally-doped LSCO. [52]

This interpretation evolved considerably in recent years as a result of further measurements and calculations. Pintschovius and Braden repeated the experiment using different experimental conditions, which had a higher energy resolution due to their use of the Cu220 monochromator. [53] They also measured the interesting wavevector range between $\mathbf{q}$=0.1 and 0.4 in the so-called focusing condition with the tilt of the resolution ellipsoid matching the phonon dispersion, which further improved the resolution compared to Ref. [52] (see sec. 4.2 for a more detailed discussion) Pintschovius and Braden reported enhanced linewidth near the same wavevector (with the strongest broadening at $\mathbf{q}$=(0.3, 0, 0)), but did not see any splitting of the phonon line. (Fig. 13)

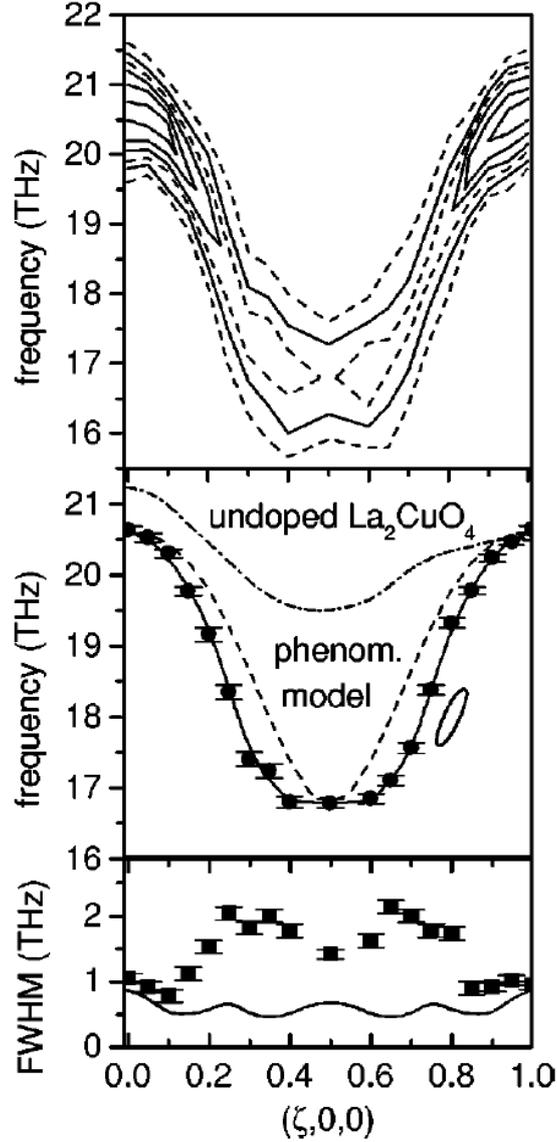

Figure 13. Results of measurements on La$_{1.85}$Sr$_{0.15}$CuO$_4$ performed with a better energy resolution and similar in-plane wavevector resolution than in Fig. 12. (from [53])

The origin of these effects was not clear at the time, but phonon anomalies near half way to the zone boundary, where static charge stripes appear in some cuprates, suggested a possible connection to incipient stripe formation.

D. Reznik et al. investigated the same bond stretching branch in La$_{1.875}$Ba$_{0.125}$CuO$_4$ where static stripes appear at low temperatures. [54] They performed the first set of measurements in the same scattering geometry as Pintschovius and Braden [53] (at wavevectors (5-4.5,0,0) or (5-4.5,0,1)) and found that the phonon dispersion could be described with two components (see Fig. 14): One had a "normal" dispersion following a cosine function (blue line), and the other softened and broadened abruptly at $\mathbf{q}$=(0.25, 0, 0) (black line). The possible relationship between the phonon anomaly and stripe formation is further explored in Sec. 4.5.

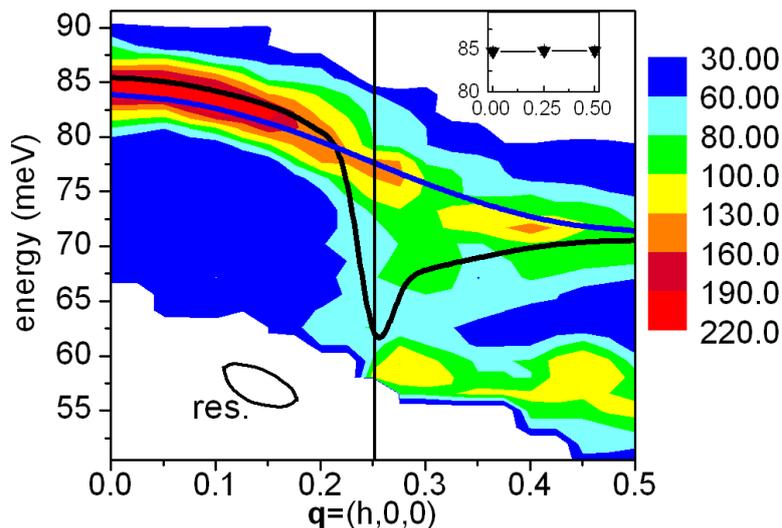

Figure 14. Color-coded contour plot of the phonon spectra observed on $La_{1.875}Ba_{0.125}CuO_4$ at 10 K. The intensities above and below 60 meV are associated with plane-polarized Cu-O bond-stretching vibrations and bond-bending vibrations, respectively. Lines are dispersion curves based on two-peak fits to the data. The white area at the lower left corner of the diagram was not accessible in this experiment. The ellipse illustrates the instrumental resolution. The inset shows the dispersion in the [110]-direction. The vertical line represents the charge stripe ordering wave vector. Blue/black line represents the "normal"/"anomalous" component respectively in the original interpretation of the authors. Subsequent work showed that a substantial part of the intensity in the "normal" component is an artifact of finite wavevector resolution in the transverse-direction (out of the page).

D. Reznik et al. [54,55] found that there was an overall hardening of the spectral weight on heating in both $La_{1.875}Ba_{0.125}CuO_4$ and $La_{1.85}Sr_{0.15}CuO_4$ at $\mathbf{q}$=(0.25,0,0). (see Fig. 15) This indicates that the anomalous broadening does not originate from anharmonicity or structural inhomogeneity, since these have the opposite or no temperature dependence.

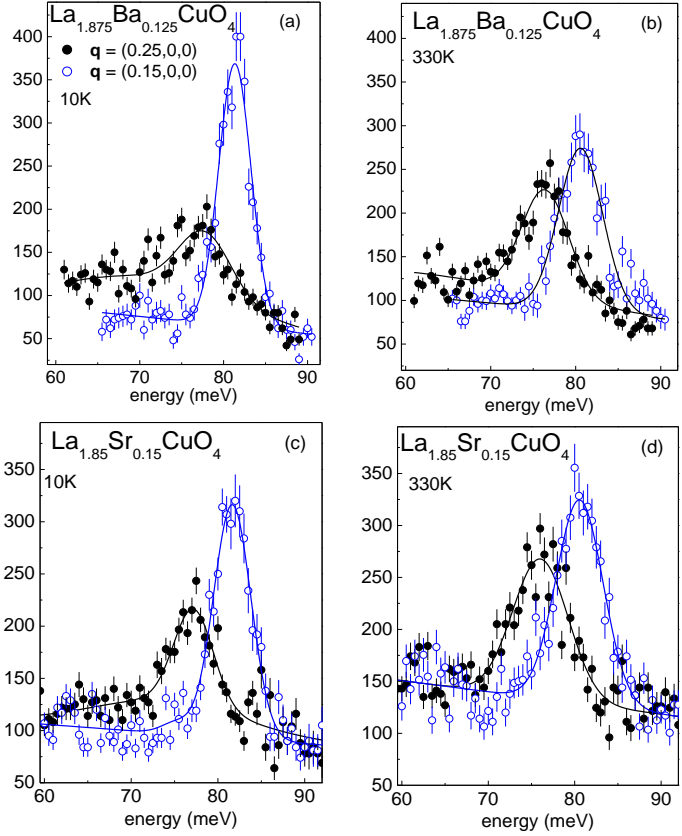

Figure 15. Temperature dependence of the bond-stretching phonons at select wavevectors. Energy scans taken on $La_{1.875}Ba_{0.125}CuO_4$ (a,b) and on $La_{1.85}Sr_{0.15}CuO_4$ (c,d,e) at 10K (a,c) and 330K (b,d) (Ref. 55). The phonon at $\mathbf{q}$=(0.15,0,0) is "normal" in that it has a Gaussian lineshape on top of a linear background. This background results from multiphonon and incoherent scattering and has no strong dependence on $\mathbf{Q}$. The intensity reduction of this phonon in $La_{1.875}Ba_{0.125}CuO_4$ from 10K (a) to 330K (b) is consistent with the Debye-Waller factor. At $\mathbf{q}$=(0.25,0,0), there is extra intensity on top of the background in the tail of the main peak. It originates from one-phonon scattering that extends to the lowest investigated energies, while the peak intensity is greatly suppressed as discussed in the text. The effect is reduced but does not disappear at 330K. Note that 330K in (b) is shown instead of 300K in the same plot in Ref. 54 because of a typographical error in the latter. Integrated intensity of the phonon decreases from $\mathbf{q}$=(0.15,0,0) to $\mathbf{q}$=(0.25,0,0) due to the decrease of the structure factor. (from [55])

Pintschovius and Braden [53] observed no temperature dependence at $\mathbf{q}$=(0.3,0,0), (it is equivalent to $\mathbf{q}$=(0.7, 0, 0) if interlayer interactions are neglected) although the zone center phonon of the same branch softened on heating. This softening is due to increased anharmonicity and should affect the entire branch. The absence of softening at $\mathbf{q}$=(0.3,0,0) that they report, indicates that there is a counterbalancing trend, which makes their results agree qualitatively with Refs. [54,55]. I will explain the reason for the quantitative difference in the following section.

McQueeney et al. [52] reported the suppression of the anomalous behavior at room temperature. However, the room temperature data of [52] suffered from a much stronger background than the low temperature data and relatively large statistical error. Ref. 55, which had a much better resolution and signal-to-background ratio, but was limited to only three wavevectors, also reported a suppression of the anomalous behavior at 330K. In this regard the two studies are consistent, although Ref. 52 claims a much more radical change of the phonon dispersion than reported in Ref. 55. To resolve this disagreement it is necessary to perform measurements covering the entire BZ at 300K with the experimental configuration of Ref. 55.

### 4.2 Recent IXS Results.

Neutron scattering experiments have a relatively poor **Q** resolution. For the cuprates its full width half maximum (FWHM) is on the order of 15% of the in-plane Brillouin zone. The effects of finite **Q** resolution in the longitudinal direction have been carefully considered in early studies, but the finite resolution in the transverse direction has not. In this section I will discuss recent IXS work and will show that some previous experiments need to be reinterpreted taking into account the finite transverse **Q**.

More recent measurements using both INS [55] and IXS [56] yielded a somewhat surprising result that the anomalous softening/broadening for **q**=(0.25,k,0) occurs only very close to k=0. For example in $La_{1.84}Nd_{0.04}Sr_{0.12}CuO_4$ the phonon anomaly significantly weakened at |k|≈0.08 compared to k=0, and disappeared entirely at |k|=0.16 (see Fig. 16). [56] The "normal" component for k≈0 was significantly suppressed.

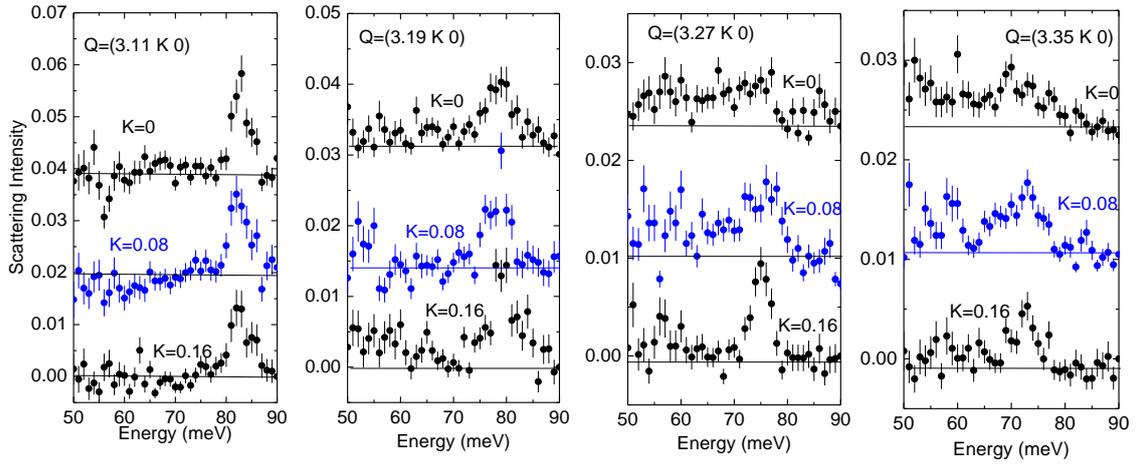

Figure 16. IXS energy scans after subtraction of the elastic tail and a constant term corresponding to the stray radiation. The scans were taken with **q**=(h,k,0) with h = 0.11, 0.19, 0.27, 0.35 (from the left to the right column) and k as indicated in the figure. The most interesting features are the suppression of the two-component behavior seen by INS at 77 meV near **Q**=(3.27,0,0) and the rapid narrowing and hardening of the phonon line from k=0 to k=0.16 for **Q**=(3.27,k,0). From Ref. [56]

Neutron measurements have a much lower resolution in the k-direction, i.e. INS experiments nominally performed with k=0 include a significant contribution from wavevectors with |k|>0.08 even in the most optimal configuration (a-b scattering plane). Thus most of the intensity in the "normal" component in the INS measurements probably comes from these phonons with |k|>0.08.

With this information it now becomes possible to explain why the temperature effect in [53] was weaker than in [52]. The experiment of Pintschovius and Braden [53] was performed in the a-c scattering plane, which had poorer wavevector resolution in the k-direction, whereas the other study was performed in the a-b scattering plane, which had a better k-resolution. [57] Since the phonon anomaly is sharp in the k-direction, the anomalous behavior should be masked by the "normal" phonons with |k|>0 in the a-c scattering plane more than in the a-b scattering plane. This "masking" would also reduce the phonon linewidth in [53] compared to the measurement in [57] performed with better k-resolution.

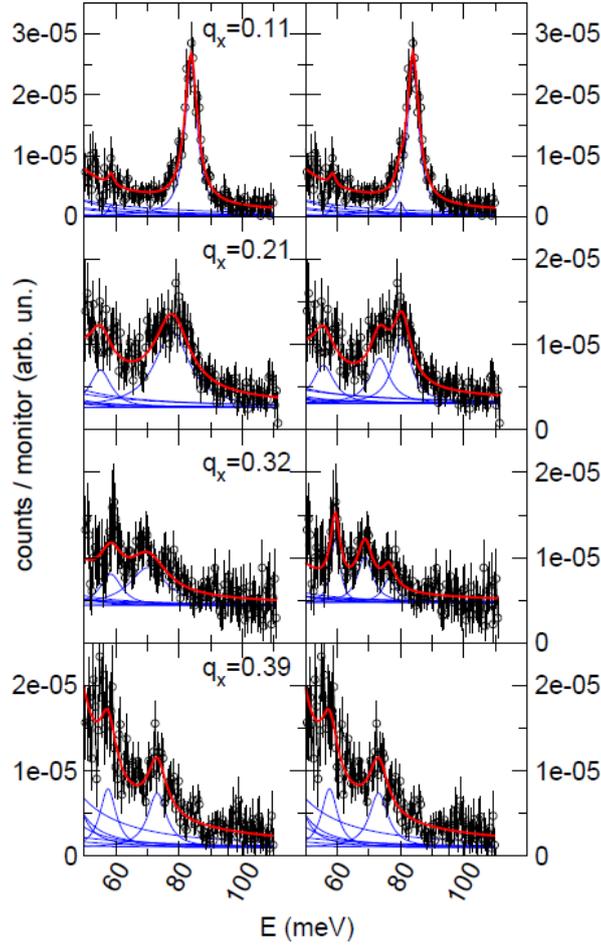

Figure 17: Inelastic X-ray scattering spectra of $La_{1.86}Ba_{0.14}CuO_{4+\delta}$, at $\mathbf{Q}$= (3+$q_x$, $q_y$, 0) ($q_y$<0.04). In the left column a single peak is used to fit the data. In the right column the data are fitted with two Cu-O bond stretching modes. (from Ref. [58])

D'Astuto et al. [58] reported the two-branch behavior in the IXS spectra of $La_{1.86}Ba_{0.14}CuO_4$ with clearly resolved "normal" and "anomalous" components. (Fig. 17) This result seems to be different from the $\mathbf{Q}$=(3.27, 0, 0) data of Fig. 16 measured also by IXS with a much higher energy resolution, as well as with the results of J. Graf et al. on $La_{1.92}Sr_{0.08}CuO_4$ [59] and Sasagawa et al. [60], which are consistent with either a single broad peak or two strongly overlapping peaks. It is possible that since D'Astuto et al. measured a Ba-doped sample, and D. Reznik et al. investigated the Sr-doped systems, the difference may come from Ba vs. Sr doping. It is necessary to perform further experiments to clarify this potentially important issue as discussed by De Fillips et al. [61]

### 4.3 Doping dependence

Figure 18 shows the bond-stretching phonon dispersion and linewidth for $La_{2-x}Sr_xCuO_4$ with x=0.07,0.15, and 0.3. The dispersion is compared with the cosine function that typically comes out of DFT/LDA/GGA calculations (see for example [62,63,64]). Here the data presented in Fig. 4 of Ref. 54 are combined with some unpublished results and refitted using the model that includes all phonon branches picked up by the spectrometer resolution as opposed to gaussian peaks. [65] Such an analysis provides more accurate values of intrinsic phonon linewidths. The strongest dip below the cosine function and the biggest peak of the linewidth is observed at optimal doping where the $T_c$ is highest. These are smaller at x=0.07 and disappear in the overdoped nonsuperconducting sample with x=0.3. The position of the phonon anomaly does not change with doping.

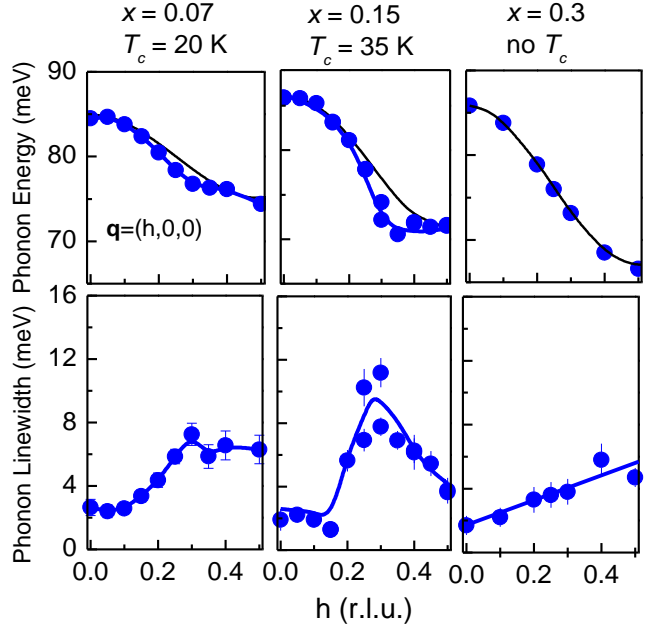

Figure 18. Bond-stretching phonon dispersion (top row) and linewidth (bottom row) in La$_{2-x}$Sr$_x$CuO$_4$ at three doping levels. Black lines represent downward cosine dispersion. The overall increase of the bond-stretching mode linewidth towards the zone boundary appears to be doping-independent. Softening compared with the cosine dispersion as well as the linewidth enhancement half-way to the zone boundary do not shift with h between x=0.07 and 0.15. Note that the enhancement of the linewidth at x=0.07 near the zone boundary compared with the higher dopings is unrelated to the anomaly at h=0.3 and can be explained by the inhomogeneous doping effect discussed in sec. 4.1 and [51].

Comparison with the overdoped sample, where the physics are conventional, allows to identify the effects of electron-phonon coupling that are intrinsic to optimal doping. Figure 19 shows the schematic of the anomalous phonon broadening that appears on top of the broadening observed in the x=0.3 sample. The anomalous broadening peaks at **q**=(0.3,0,0) and weakens rapidly in the longitudinal and transverse directions.

This effect is phenomenologically very similar to the renormalization of the acoustic phonons at specific wavevectors discussed in sections 2 and 3. Next, I will show that profound differences exist between La$_{2-x}$Sr$_x$CuO$_4$ and conventional metals.

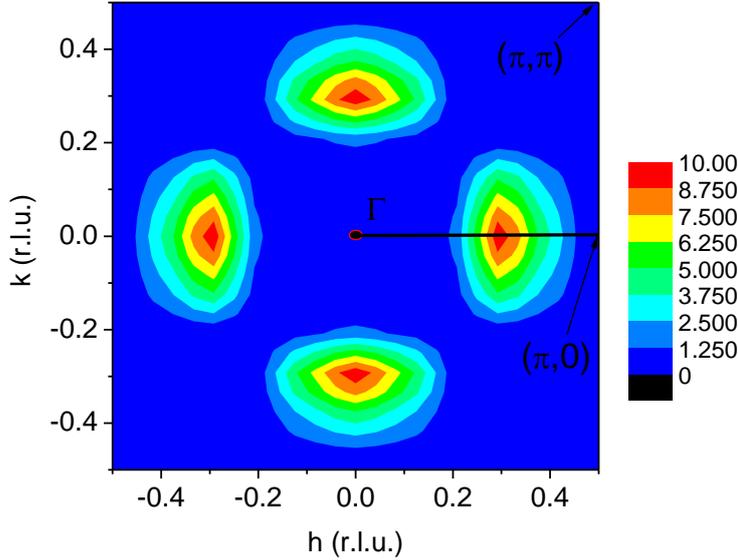

Figure 19. Qualitative picture of the difference in the linewidths of the bond-stretching phonon in optimally-doped (x=0.15) and overdoped (x=0.3) $La_{2-x}Sr_xCuO_4$ as a function of wavevector in the ab-basal plane based on [50] and [56]. The units of the color scheme are meV. The solid line indicates the [100] direction along which most of measurements were performed.

### 4.4 Comparison with density functional theory

As discussed in Sec. 2, density functional theory gives a good description of phonon dispersions in metals where electron-electron interactions beyond the mean field level (as taken into account in LDA and GGA) can be neglected. Giustino et al. [63] performed such a calculation in the GGA for $La_{1.85}Sr_{0.15}CuO_4$, which approximately agreed with the early experimental data of [52]. In particular, they reproduced the overall downward dispersion of the longitudinal bond-stretching branch. However, the strong effect in the bond-stretching phonon half way to the zone boundary was not apparent in figure 1 of Ref. [63], because the 300K results and 10K results were plotted together. In the brief communiation arising from the article of Giustino et al, D. Reznik et al. showed that DFT did not reproduce the phonon anomaly half way to the zone boundary that appears in Ref. [50,52,53,54,55,56,58,59] as well as in later experiments discussed above in detail. (Fig. 20, [66]).

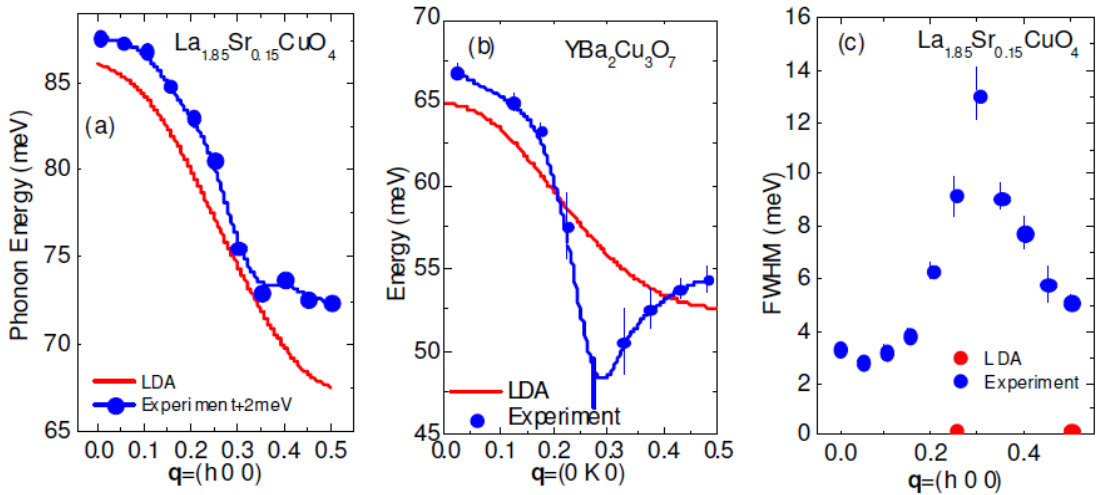

Figure 20. Comparison of some LDA predictions with experimental results for $La_{1.85}Sr_{0.15}CuO_4$ [54] and $YBa_2Cu_3O_7$ [86] at 10K. (a,b) Experimental bond-stretching phonon dispersions compared to LDA results. The data in (a) are shifted by 2meV. (c) Phonon linewidths in $La_{1.85}Sr_{0.15}CuO_4$ compared with LDA results on $YBa_2Cu_3O_7$. Ref. [63] contains no linewidth results for $La_{1.85}Sr_{0.15}CuO_4$ but they should be similar.

It is interesting that many-body calculations predict a substantial enhancement of the coupling to bond-stretching phonons compared to DFT. (see for example refs. 49 and 67) t-J model-based calculations describe interesting doping dependence of the zone boundary phonons, suggesting that strong correlations might be relevant. [68,69,70] The idea that electronic correlations are responsible for the enhanced electron-phonon coupling is further reinforced on the qualitative level by theoretical investigations based on the Hubbard-Holstein model and similar models, which find an enhancement of phonon renormalization by electronic correlations not included in LDA. [71]

Strong renormalization of the bond stretching phonons has been taken as evidence for a soft collective charge mode [72,73] or an incipient instability [74] with respect to the formation of either polarons, biporarons [75,76,77], charge density wave order [78], phase separation [79,80,81], valence bond order [82], or other inhomogeneity [83]. These may or may not be related to the mechanism of stripe formation. A number of studies suggested that these instabilities may lead to superconductivity. [74,75,76,80]

Further evidence that the bond-stretching phonon anomaly results from electronic correlations comes from an excellent agreement between the GGA calculation and the experimental phonon dispersion in nonsuperconducting overdoped $La_{1.7}Sr_{0.3}CuO_4$. [50, Fig. 18 in Ref. 11]

### 4.5 Connection with stripes and other charge-inhomogeneous models

The bond-stretching phonon anomaly is strongest in $La_{1.875}Ba_{0.125}CuO_4$ and $La_{1.48}Nd_{0.4}Sr_{0.12}CuO_4$, compounds that exhibit spatially modulated charge and magnetic order, often called stripe order. It appears when holes doped into copper-oxygen planes segregate into lines, which act as domain walls for an antiferromagnetically ordered background. Static long-range stripe order has been observed only in a few special compounds such as $La_{1.48}Nd_{0.4}Sr_{0.12}CuO_4$ and $La_{1.875}Ba_{0.125}CuO_4$ where anisotropy due to the transition to the low temperature tetragonal structure provides the pinning for the stripes while superconductivity is greatly suppressed. [4] In contrast, the more common low temperature orthorhombic (LTO) phase does not provide such a pinning and static stripes do not form. In the LTO phase the stripes are assumed to be purely dynamic, which makes their detection extremely difficult. [5] Here I discuss the possible relation between the phonon anomaly and dynamic stripes.

A detailed comparison between the bond-stretching phonon dispersion in stripe-ordered compounds and optimally-doped superconducting $La_{1.85}Sr_{0.15}CuO_4$ was performed by D. Reznik et al. [55] They found the strongest phonon renormalization at h=0.25 in the presence of static stripes and h=0.3 at optimal doping. (Fig. 21) It appears that static stripes pin the phonon anomaly at the stripe ordering wavevector.

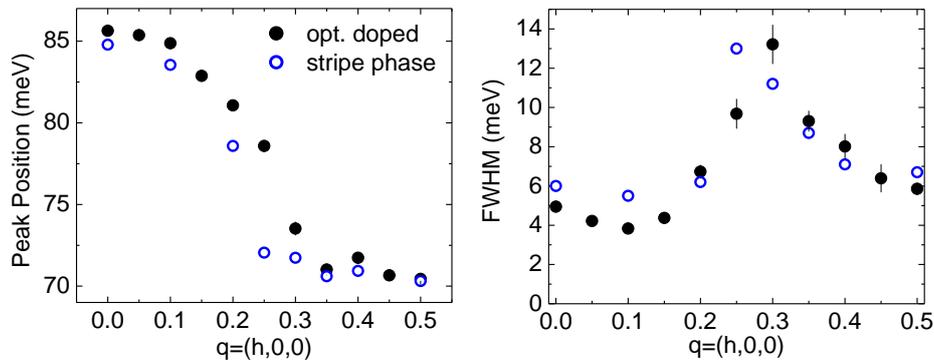

Figure 21. Comparison of the phonon dispersions (a) and linewidth (b) of the bond-stretching branch in $La_{1.85}Sr_{0.15}CuO_4$ and $La_{1.48}Nd_{0.4}Sr_{0.12}CuO_4$. (from Ref. [55])

Two mechanisms of the impact of dynamic stripes on phonons have been proposed: One is that the phonon eigenvector resonates with the charge component of the stripes; The other is that one-dimensional nature of charge stripes makes them prone to a Kohn anomaly, which renormalizes the phonons. In the first scenario (2D picture) the propagation vector of the anomalous phonon must be parallel to the charge ordering wavevector, whereas in the second scenario (1D picture) it must be perpendicular to the charge ordering wavevector. (Fig. 22)

An important clue is that the phonon anomaly disappears quickly as one moves away from k=0 along the line in reciprocal space: $\mathbf{q}$=(0.25,k,0) as shown in Refs. [55] and [56] and discussed in section 2.2. Such behavior is expected from the matching of the phonon wavevector and the stripe propagation vector. In contrast, a simple picture of a Kohn anomaly due to 1-D physics inside the stripes predicts a phonon anomaly that only weakly depends on k. This observation favors the 2D picture, but an important caveat is that it may be possible to reconcile the 1D picture with experiment by including a decrease of the electron-phonon matrix element away from k=0. [84]

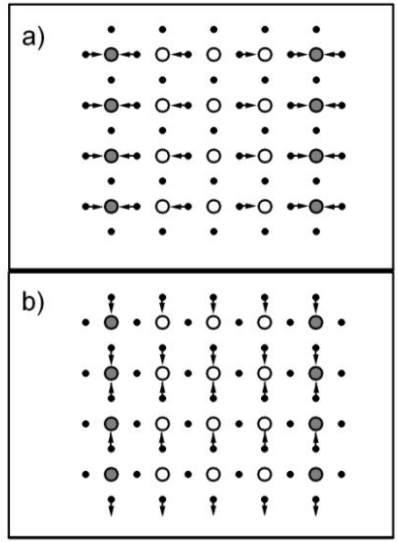

Figure 22. Schematic of the eigenvectors for the phonons with $\mathbf{q}$=(0.25 0 0) propagating perpendicular (a) and parallel (b) to the stripes. Open circles represent hole-poor antiferromagnetic regions, while the filled circles represent the hole-rich lines. (from Ref. [56])

Another way to distinguish between the two scenarios is to consider the doping dependence of the wavevector of the maximum phonon renormalization, $\mathbf{q}_{max}$. In the stripe picture, $\mathbf{q}_{max} = 2\mathbf{q}_{in}$, ($\mathbf{q}_{in}$ is the wavevector of incommensurability of low energy spin fluctuations). [5] At doping levels of x=0.12 and higher, $\mathbf{q}_{in}$= 0.125, which gives the charge ordering wavevector of 0.25. This value is indeed close to $\mathbf{q}_{max}$. At x=0.07 $\mathbf{q}_{in}$= 0.07. This gives the charge stripe ordering wavevector of 0.14 whereas $\mathbf{q}_{max}$=0.3. This discrepancy appears to contradict the 2D picture. But again there is a caveat: Anomalous phonons occur at a fairly high energy of about 75 meV, and a comparison to the dynamic stripe wavevector at low energies may not be appropriate.

Thus the question of which picture, 1D or 2D, agrees better with the data is not yet settled.

If the phonon renormalization is driven by static stripes, one may expect to see different behavior for phonons propagating parallel or perpendicular to the stripe propagation vector. [43] In this case the phonon should split into two peaks. The dynamic stripes, according to M. Vojta et al. [85], may not break tetragonal symmetry, because fluctuations can occur in both directions simultaneously. Thus a single-peak anomalous phonon lineshape is compatible with dynamic stripes.

## 5. Other Cuprates
### 5.1 Bond-Stretching phonon anomalies in YBa$_2$Cu$_3$O$_{6+x}$

It is necessary to establish the universality of the phonon anomalies observed in the La$_2$-$_x$Sr$_x$CuO$_4$ family. Until now much less work has been performed on other cuprates, because they are more difficult to measure either due to the higher background, no availability of large samples for INS, or low IXS scattering cross section.

In the case of YBa$_2$Cu$_3$O$_{6+x}$ the orthorhombic structure combined with twinning complicates the interpretation of the results. Very little work has been done so far on detwinned samples [86] because they are smaller than the twinned ones. Furthermore, two CuO$_2$ layers in the unit cell introduce two bond-stretching branches, of $\Delta$1 and $\Delta$4 symmetry.

At optimal doping, bond-stretching phonons propagating along the chain direction show an anomaly that is in many respects similar to the one in La$_2$-$_x$Sr$_x$CuO$_4$. [87,88] It is absent at 300K, and appears at low temperatures. Chung et al. [88] reported that the spectral weight of the bond-stretching phonons in the $\Delta$1 symmetry redistributes to lower energies below the superconducting transition temperature, T$_c$=93K. (see Fig. 23)

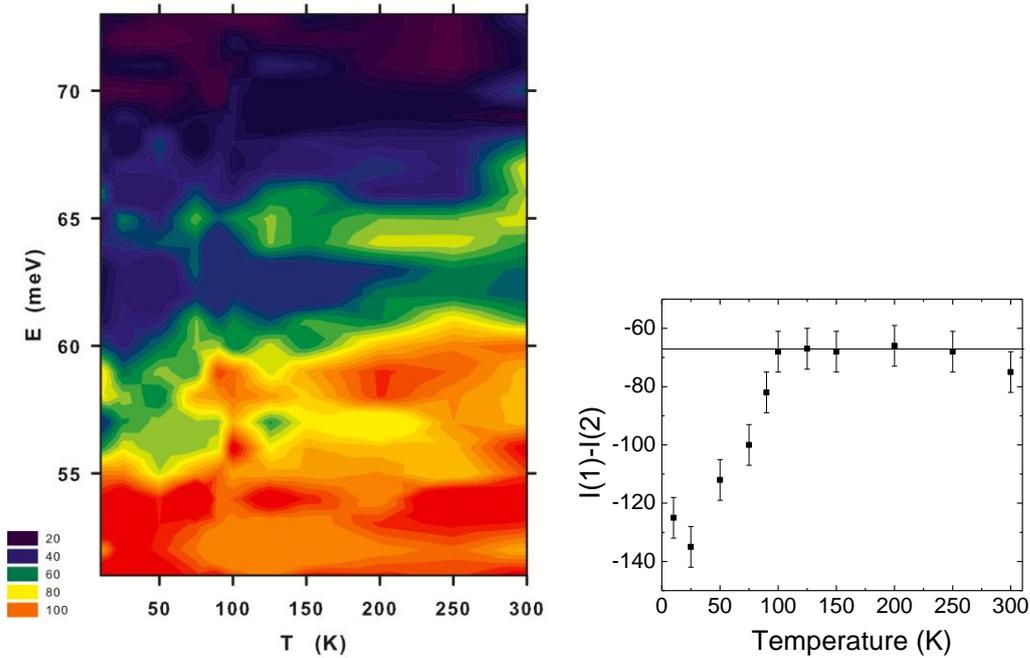

Figure 23. (right) Inelastic scattering intensity of YBa$_2$Cu$_3$O$_{6.95}$ at Q=(3.25,0,0) as a function of temperature, determined with the triple-axis spectrometer at the HFIR. Data were smoothed once to reduce noise. (left) Temperature dependence of the intensity difference (I1)-(I2), where I1 is the average intensity from 56 to 68 meV, I2 from 51 to 55 eV, at **Q**=(3.25,0,0). T$_c$ of the sample was 93K. (from Ref. [88])

L. Pintschovius et al. [87] and D. Reznik et al. [89] found that a similar transfer of spectral weight occurs for the $\Delta$4 phonons but starting close to 200K, not at T$_c$. (Fig. 24a) They interpreted this transfer of spectral weight as arising from softening of the bond-stretching phonon polarized along b*, which transfers its eigenvector to the branches that are lower in energy. This interpretation could explain the observed behavior with some important caveats, but more work is necessary to better understand this effect. Figure 24b shows that this transfer of spectral weight accelerates below T$_c$ saturating near 50K. While clearly related to the onset of superconductivity, this effect is not understood.

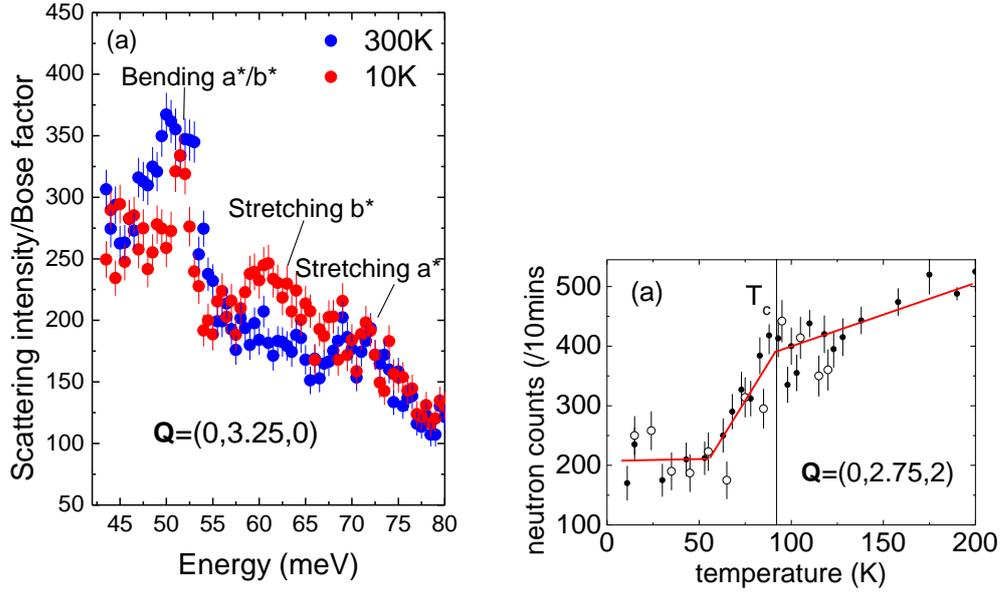

Figure 24. Data of Ref. [89] for the $\Delta 4$ symmetry in $YBa_2Cu_3O_{6.95}$. (a) Comparion of the 300K and 10K spectra. (b) Backround-subtracted intensity at Q = (0, 2.75,−2) and E = 60 meV (see text). Open and solid circles represent different datasets.

D. Reznik et al. also showed that the transfer of spectral weight in the $\Delta 1$ symmetry also begins well above $T_c$ with the most pronounced change below $T_c$. (Fig. 25) This result seems to contradict the observation of Chung et al. [88] who reported that the phonon intensity shift in $\Delta 1$ symmetry occurs only below $T_c$. According to Reznik et al. the effect would also appear only below $T_c$ if they excluded the intensity below 50 meV from their analysis [89] as was done in Ref. [88]. So in this respect the two studies are consistent.

The phonon anomaly in $YBa_2Cu_3O_{6.95}$ seems to extend far in the transverse direction (Fig. 26), i.e. it may be consistent with the 1D picture. [89] (also see sec. 4.5) However, twinning of the sample made the data difficult to interpret. Otherwise, the phonon anomaly in optimally-doped $YBa_2Cu_3O_{6.95}$ is similar to the effect in $La_{2-x}Sr_xCuO_4$.

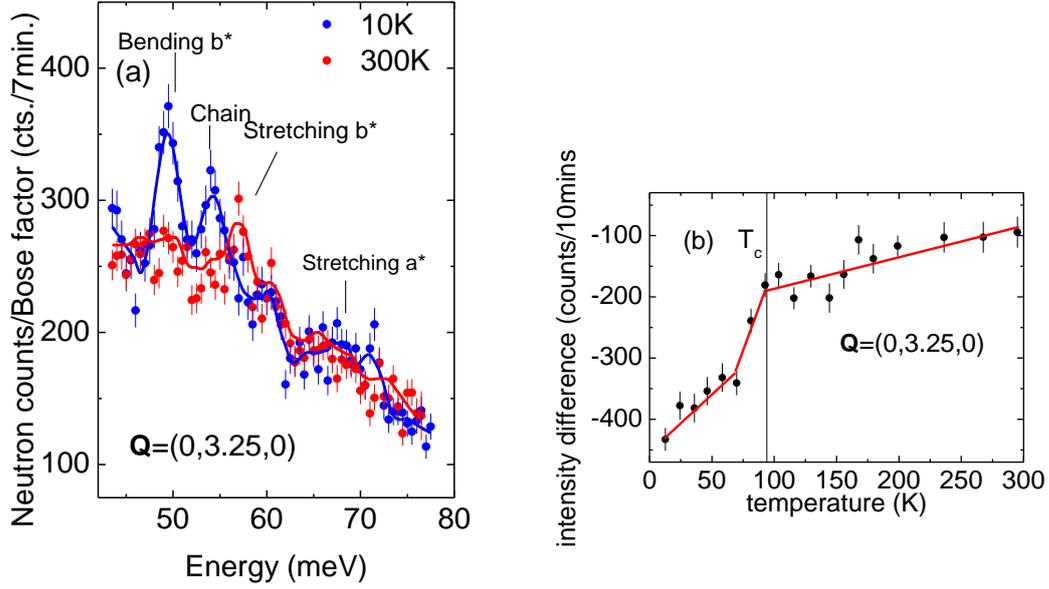

Figure 25. Data of Ref. [89] for the Δ1 symmetry. (a) Phonon spectra at 300K and 10K. The main difference with Ref. [88] and Fig. 23 is a bigger energy range here: 42-75meV in [89] vs. 51-72 meV in [88] (b) The difference between the intensity at 57 meV and the average of intensities at 53 meV and 49 meV at $\mathbf{Q} = (0, 3.25, 0)$. Temperature dependence above $T_c$ not seen in Fig. 23 comes from including the 49meV phonon, which falls outside the energy range investigated in [88].

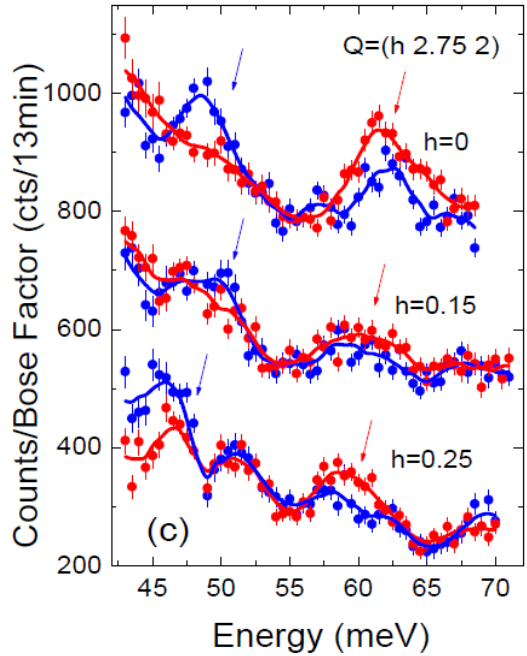

Figure 26. Energy (E) scans taken at 200K (red) and 10K (blue). Data were taken with the final energy $E_f = 13.4$ meV for E > 46 meV and with $E_f = 12.5$ meV for 43 <E <48 meV. The 12.5-meV data were corrected for the different resolution volume by multiplying by $(13.4/12.5)^2$. The resulting intensities were averaged in the overlapping energy range (46.5-48meV). The 200K data were divided by the Bose factor and 23 counts were subtracted to correct for the temperature dependence of the background. Blue/red arrows indicate intensity gain/loss. (from Ref. [89])

Much less is known about YBa$_2$Cu$_3$O$_{6+x}$ at lower doping levels. Stretzel et al [90] reported splitting of the bond-streching branch arguing in favor of charge inhomogeneity, whereas Pintschovius et al. [91] explained similar results in terms of the difference between the dispersion of the stretching phonons propagating parallel and perpendicular to Cu-O chains. This disagreement needs to be settled by measurements on detwinned samples.

### 5.2 Bond-Buckling phonon anomalies in YBa$_2$Cu$_3$O$_{6+x}$

Early Raman scattering experiments on optimally doped cuprates have shown that the bond-buckling mode exhibits a superconductivity-induced softening of ~1.5% at the wave vector **q** = 0. [92] This work was followed by neutron scattering measurements on a large twinned single crystal, which showed that the softening persists at nozero **q** in the directions along the Cu-O bonds (100 and 010-directions also called a* and b* respectively), but not in the 110 direction. [93] This result was interpreted in terms of the interaction of the phonon branch with the d-wave superconducting gap. Recently M. Raichle et al. [94] investigated a detwinned sample and found that the softening of the phonon decreased away from the zone center along a*, but showed a pronounced enhancement along b* around k=0.3. (Fig. 27)

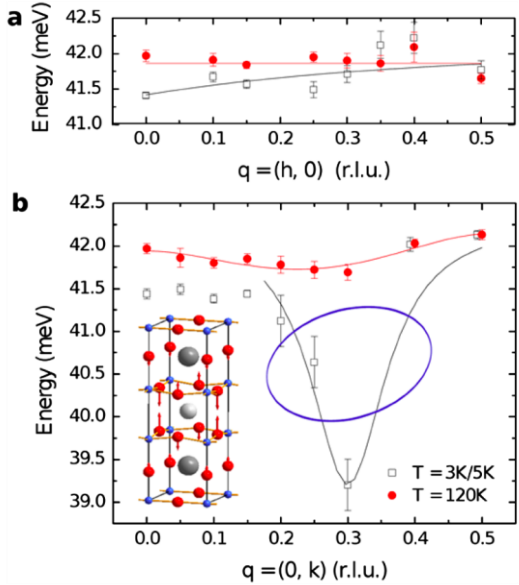

Figure 27. Dispersion of the buckling mode in optimally-doped YBa$_2$Cu$_3$O$_{6+x}$ along the a* and b*. The black line is the dispersion used for the resolution calculation. A projection of the four-dimensional resolution ellipsoid is shown for comparison. The data point at in-plane **q**$_{in}$ = (0, 0.3) is the result of the resolution convolution; the remaining points were determined by fits to standard Voigt functions. The inset shows the eigenvector of the buckling mode at **q**$_{in}$ = (0, 0.3). The elongations of the apical oxygen atoms and of the in-plane oxygen atoms along b were enlarged by a factor of 4 for clarity. [94]

M. Raichle et al. concluded that the anomalous phonon with such a large a-b anisotropy cannot mediate d-wave pairing, but may contribute to the gap anisotropy and the hotly-debated kink in the electronic dispersion observed by angle-resolved photoemission. More importantly, both the bond-stretching and bond-buckling modes indicate that there is a charge fluctuation in YBa$_2$Cu$_3$O$_{6+x}$ with the wavevector close to **q**=(0.3,0,0), which indicates nematic behavior. It is possibly associated with stripe formation with stripes running along the a-axis (perpendicular to the chains). An alternative explanation also proposed in [94] is that it could be related to CDW-type instability in the copper oxygen chains.

### 5.3 HgBa$_2$CuO$_{4+x}$

Bond-stretching phonons in HgBa$_2$CuO$_{4+x}$ have been measured by Uchiyama et al. [95] These measurements showed that the bond-stretching phonons soften similarly to La$_{2-x}$Sr$_x$CuO$_4$ and YBa$_2$Cu$_3$O$_{6+x}$. (Fig. 28) It is important to extend this study to different dopings, temperatures and nonzero transverse wavevectors.

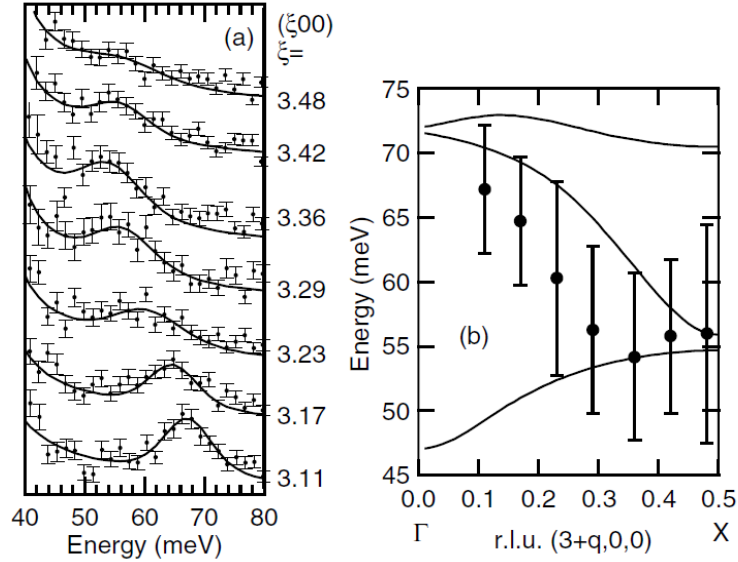

Figure 28. Bond stretching phonons in $HgBa_2CuO_{4+x}$. (a) Enlarged spectra taken close to the bond stretching mode plotted on a linear scale. (b) Data points represent frequencies of the bond-stretching phonons. The lines show the shell model calculation in which the interaction between the next-nearest neighbor oxygens in the $CuO_2$ plane is added. The lines indicate (top to bottom) the $c$-polarized apical oxygen mode, the $a$-polarized Cu-O bond stretching mode, and the $a$-polarized in-plane Cu-O bending mode, respectively. The vertical bars indicate the FWHM of the peaks determined in fitting data shown in (a). (from [95])

### 5.4 $Bi_2Sr_{1.6}La_{0.4}Cu_2O_{6+x}$

Graf et al. [96] measured phonon dispersions by IXS and electronic dispersions by ARPES in a single-layer Bi-based cuprate, $Bi_2Sr_{1.6}La_{0.4}Cu_2O_{6+x}$. They reported a similar phonon anomaly as in other cuprates (Fig. 29) and argued in favor of a correlation between this phonon anomaly, the kink observed in photoemission and the Fermi arc that characterizes the pseudogap phase. They related the sudden onset of phonon broadening near $\mathbf{q}$=(0.2,0,0) to coupling of the phonon to the Fermi arc region of the Fermi surface, but not to the pseudogap region. The Fermi arc region is not nested, so exceptionally large electron-phonon coupling for the stretching branch is necessary for this interpretation to be valid. (as in Cr as described in Sec. 3.6) In addition one needs to consider that in other families of cuprates, where the doping dependence has been investigated, the wavevector of the onset of the phonon effect is doping independent, whereas the length of the Fermi arc strongly depends on doping. More detailed studies of this compound, especially as a function of doping, are necessary to clarify these issues.

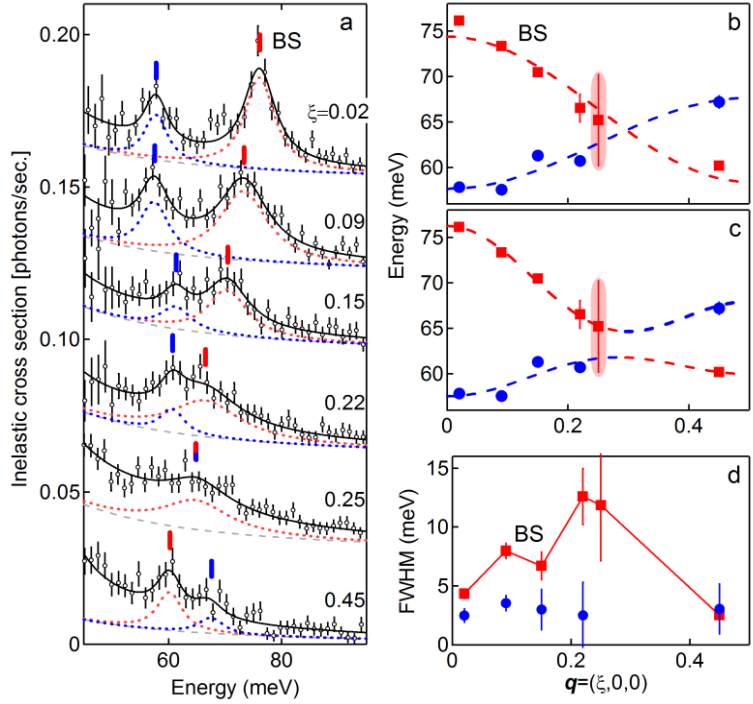

Figure 29. LO phonon dispersions in Bi$_2$Sr$_{1.6}$La$_{0.4}$Cu$_2$O$_{6+x}$. [96] (a) IXS spectra for Q=(3+ξ,0,0) with ξ from the BZ center (top spectrum, ξ = 0.02) to the BZ boundary (bottom spectrum, ξ = 0.45). The spectra are vertically shifted. The solid lines show the harmonic oscillator fit, the dashed lines show the elastic tail and the dotted lines show the two modes used in the fit. (b,c) Phonon dispersions and linewidths. The cosine dashed lines are guides for the eyes illustrating the crossing (b) and anticrossing (c) scenarios. (d) Full width at half maximum. The error bars are an estimate of the standard deviation of the fit coefficients.

## 5.5 Electron-doped cuprates

Bond-stretching phonons have been investigated in electron-doped cuprates only in Nd$_{2-x}$Ce$_x$CuO$_4$. Phonon density of states measurements on powder samples showed that electron doping softens the highest energy oxygen phonons as occurs in the case of hole-doping. [97] The first single crystal experiment has been performed by d'Astuto et al. by IXS [98] who found that the bond-stretching phonon branch dispersed steeply downwards beyond h=0.15. This work was, in fact, the first IXS experiment on the high T$_c$ cuprates. These measurements, however were complicated by the anticrossing of the bond-stretching branch with another branch due to Nd-O vibrations that dispersed sharply upwards. The anticrossing occurs near h=0.2 complicating the interpretation of the data near these wavevectors. Another difficulty came from low IXS scattering cross sections for the oxygen vibrations.

A neutron scattering investigation has been performed by M. Braden et al. [99] once large single crystals became available. Oxygen phonons have a higher scattering cross section in the INS than in the IXS experiments, allowing a more accurate determination of the phonon dispersions.

The two studies showed that the bond-stretching phonon dispersion in Nd$_{1.85}$Ce$_{0.15}$CuO$_4$ was similar to that in the hole-doped compounds. (Fig. 30) This similarity points at a commonality between the tendencies to charge inhomogeneity between the hole-doped and electron-doped compounds as discussed in Ref. [99].

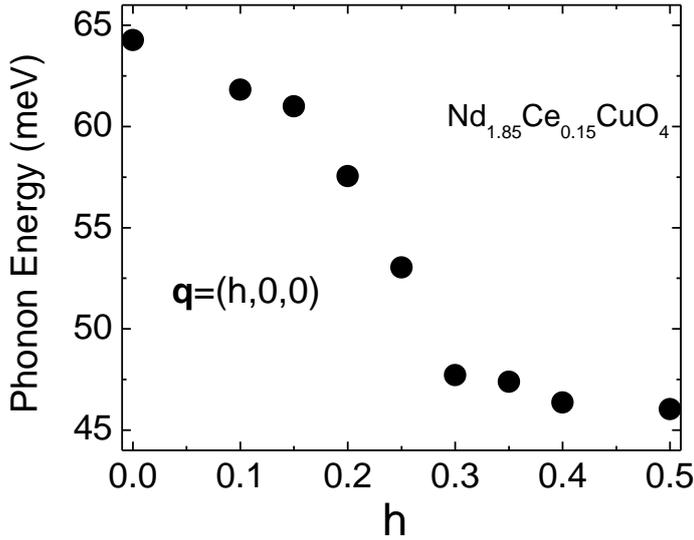

Figure 30. Dispersion of the Cu-O bond-stretching phonon in $Nd_{1.85}Ce_{0.15}CuO_4$ adapted from Ref. [99]

## 6. Conclusions

A lot of progress has been made in recent years in understanding of phonon anomalies in a variety of compounds including the ones with static or dynamic stripes. It was possible to distinguish between different mechanisms behind the phonon softening and broadening: bond-length mismatch, $\mathbf{q}$-dependence of electron-phonon coupling to conduction electrons, Fermi surface nesting, and electronic correlations.

In copper oxide superconductors, where the dynamic stripes are suspected, the bond-stretching phonons around $\mathbf{q}=(0.3,0,0)$ are softer and broader than expected from conventional theory. This effect may be related to incipient instability with respect to the formation of dynamic stripes or another charge-ordered or inhomogeneous state.

Much more experimental and theoretical work is necessary to understand the interplay between stripes and phonons. Up to now very little experimental work has been done on compounds where static stripes are clearly present, such as the nickelates and the cobaltates. Understanding the phenomenology of phonon anomalies in these compounds is essential to be able to clearly distinguish effects of stripes from other causes of phonon anomalies discussed in sections 2 and 3. In the cuprates, new evidence emerged that makes it unlikely that anomalous phonons, that may be interacting with stripes, directly mediate superconductivity (sec. 5.2). [94] However, phonon renormalization in optimally-doped YBCO accelerates at or near the superconducting $T_c$. [87,88,89,94] Furthermore, doping dependence of phonon anomalies in LSCO suggests that it is indirectly associated with the mechanism of superconductivity. Thus the relationship between phonon anomalies and the mechanism of high temperature superconductivity needs to be explored further.

## 7. Acknowledgement

Work on this article was supported by the DOE, Office of Basic Energy Sciences under Contract No. DE-SC0006939. I greatly benefited from interactions with many people over the years without which this work would not have been possible. In particular, I would like to acknowledge discussions with L. Pintschovius, R. Heid, K.-P. Bohnen, W. Reichardt, H. von Löhneysen, F. Weber, D. Lamago, A. Hamann, J.M. Tranquada, S.A. Kivelson, T. Egami, Y. Endoh, M. Arai, K. Yamada, P.B. Allen, I.I. Mazin, J. Zaanen, D.J. Singh, G. Khaliullin, B. Keimer, D.A. Neumann, J.W. Lynn, A. Mischenko, N. Nagaosa, F. Onufrieva, P. Pfeuty, P.

Bourges, Y. Sidis, O. Gunnarson, S.I. Mukhin, P. Horsch, T.P. Devereaux, Z.-X. Shen, A.Q.R. Baron, D.S. Dessau, P. Böni, M. d'Astuto, A. Lanzara, M. Greven, and M. Hoesch.